\documentclass[%
 reprint,
superscriptaddress,
 amsmath,amssymb,
 longbibliography,
 prb,
]{revtex4-2}

\usepackage[ruled]{algorithm2e}
\usepackage[noend]{algpseudocode}
\usepackage{wasysym}
\usepackage{graphicx}
\usepackage{dcolumn}
\usepackage{bm}
\DeclareMathAlphabet{\mathpzc}{OT1}{pzc}{m}{it}

\begin{document}

\preprint{APS/123-QED}

\title{Variational optimization of the amplitude of neural-network quantum many-body ground states}

\author{Jia-Qi Wang}
\affiliation{Department of Physics, Renmin University of China, Beijing 100872, China}
\affiliation{Key Laboratory of Quantum State Construction and Manipulation (Ministry of Education), Renmin University of China, Beijing 100872, China}

\author{Han-Qing Wu}
\affiliation{Guangdong Provincial Key Laboratory of Magnetoelectric Physics and Devices, School of Physics, Sun Yat-sen University, Guangzhou 510275, China}

\author{Rong-Qiang He}\email{rqhe@ruc.edu.cn}
\affiliation{Department of Physics, Renmin University of China, Beijing 100872, China}
\affiliation{Key Laboratory of Quantum State Construction and Manipulation (Ministry of Education), Renmin University of China, Beijing 100872, China}

\author{Zhong-Yi Lu}\email{zlu@ruc.edu.cn}
\affiliation{Department of Physics, Renmin University of China, Beijing 100872, China}
\affiliation{Key Laboratory of Quantum State Construction and Manipulation (Ministry of Education), Renmin University of China, Beijing 100872, China}

\date{\today}

\begin{abstract}
  Neural-network quantum states (NQSs), variationally optimized by combining traditional methods and deep learning techniques, is a new way to find quantum many-body ground states and has gradually become a competitor of traditional variational methods. However, there are still some difficulties in the optimization of NQSs, such as local minima, slow convergence, and sign structure optimization. Here, we split a quantum many-body variational wave function into a multiplication of a real-valued amplitude neural network and a sign structure, and focus on the optimization of the amplitude network while keeping the sign structure fixed. The amplitude network is a convolutional neural network (CNN) with residual blocks, namely a residual network (ResNet). Our method is tested on three typical quantum many-body systems. The obtained ground state energies are better than or comparable to those from traditional variational Monte Carlo methods and density matrix renormalization group. Surprisingly, for the frustrated Heisenberg $J_1$-$J_2$ model, our results are better than those of the complex-valued CNN in the literature, implying that the sign structure of the complex-valued NQS is difficult to optimize. We will study the optimization of the sign structure of NQSs in the future.
\end{abstract}


\maketitle


\section{Introduction}
How to obtain the ground state wave function of an interacting many-particle system has always been an important research issue in the field of quantum many-body computation, which is crucial to understanding the physical properties. However, there is no computational method that is efficient and applicable to all quantum many-body systems to date.

In recent years, with the rapid development of artificial intelligence, the related technologies have gradually been used in scientific research, and substantial achievements have been made in climate prediction~\cite{lam2023WeatherForecasting}, protein folding~\cite{2021AlphaFold}, etc. In the field of quantum many-body computation~\cite{2017Introduction,Huang2022QuantumCalc,Carleo2019Review_AI_Physics}, after the seminal attempt~\cite{GiuseppeCarleo2017_RBM} which adopted artificial neural networks as a representation of wave functions of quantum many-body systems, dubbed neural-network quantum states (NQSs), a lot of research have been done~\cite{Nomura2022ParametersRBM_NNQS,Zhang2023_LocalSequentialUpdateNNQS,roth2023_GCNN,Luciano2023_Transformer_1D_J1J2,Zhang2023Transformer_NQS_TFIM,reh2023DetailsSymmetrization,Liang2018_CNN,HibatAllah2020_RNN_NQS,Claudio2020_NQS_SignProblem,Liang2021_CNN+PEPS,Lin2022_optimize_scale,chen2023_EfficientOptimization,Nomura2017_RBM+PP,Nomura2021_RBM+PP}, such as Ferminet~\cite{Pfau2020Ferminet} for small molecular systems.

Various architectures of neural networks, such as the feedforward neural network, restricted Boltzmann machine (RBM), and convolutional neural network (CNN), have expanded the representation of the wave function of many-body systems, and their performance is comparable to that of traditional methods, such as variational Monte Carlo (VMC) methods~\cite{Sandvik2007VBBasis,Sandvik2010_VBBasis+LoopUpdate,Sorella2013_VMC,Sandvik2007_VMC+TNS,Tagliacozzo2009_TreeTensorNetwork}, density matrix renormalization group (DMRG)~\cite{GongShouShu2014_DMRG,Stoudenmire2012_2DsystemDMRG,Sandvik2018_DMRG}, and quantum Monte Carlo (QMC) methods~\cite{Sandvik1997SSEQMC,Sandvik2010_VBBasis+LoopUpdate}. This is mainly due to the improvement of algorithms and the rapid growth of GPU computing power. In the research of deep learning, the increase of network size can usually bring huge benefits, but also has expensive costs, such as large language models (LLMs)~\cite{zhao2023LLMSurvey} with emergent abilities~\cite{Wei2022EmergentAbilities}, which have aroused the interest of many researchers recently. The threshold of ``large'' in LLMs is 50 billion parameters. Its training is extremely expensive. However, increasing model size is just one way to improve the level of deep learning tasks. Other directions include improving the efficiency of optimization algorithms~\cite{ruder2017overview_GradientDescent}, improving the model architectures~\cite{vaswani2023attention,he2015Resnet}, and so on. In this work, the network architecture is redesigned and the corresponding parameters are carefully optimized, and then the potential of the neural network for quantum many-body wave function representation is explored.

In the variational optimization process of the ground state wave function of a two-dimensional frustrated system, it is hard to obtain a high-precision solution. One reason is that the sign structure of the ground state wave function for a frustrated quantum many-body system is very complex~\cite{Park2022_Frustrated_RBM,Chen2022_SignStructure-NeuralNetwork,westerhout2023SignStructure,Westerhout2020_difference_Sign_amp} and is very difficult for machine learning. In the past studies, people focused on replacing the wave function representation with a new form, namely optimizing the sign structure and amplitude of the wave function simultaneously, and expected to obtain a more accurate ground state. Here, we separate the difficulty. The wave function is split into its amplitude part and sign part. And we focus on the optimization of the amplitude, and leave the study of the sign optimization to the future.

\begin{figure*}[t!]
  \includegraphics[width=16cm]{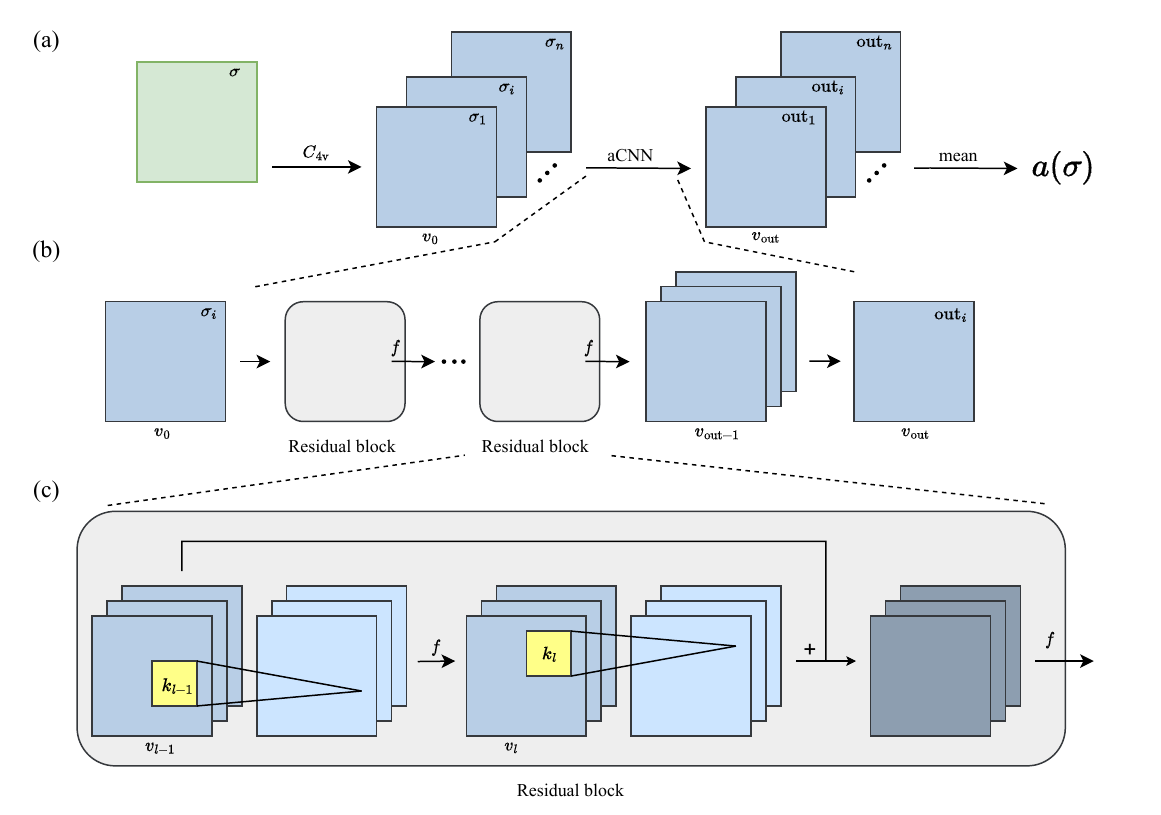}
  \caption{The architecture of the aCNN. The input is a spin configuration $\sigma$ of the system, while the output is $a(\sigma)$[Eq.~\protect\eqref{eq:amp_out}]. (a) ${\cal C}_{\rm 4v}$ symmetrization of the aCNN as a data augmentation. $v_0$ and $v_{\mathrm{out}}$ denote the input and output layers, respectively. (b) A single spin configuration $\sigma_i$ propagation in the aCNN, where $f$ is an activation function. (c) Residual block, where $k_l$ is the convolution kernel of layer $l$. The shortcut connection achieves cross-layer information transmission.}\label{fig:net}
\end{figure*}

\section{Framework}

\subsection{Neural network quantum states}

According to the universal approximation theorem~\cite{Horink1989UAT,Cybenko1989ApproximationBS}, in principle, a neural network with sufficient depth and width can fit any function. We use deep CNN to represent the amplitude of a quantum many-body wave function. In the research of deep learning, CNN has been widely used in related fields of picture processing. Compared with network architectures such as recurrent neural networks, feedforward neural networks, and RBM, CNN has remarkable advantages in extracting spatial correlation of input information.

The trial wave function is divided into two parts: sign $s(\sigma)$ and amplitude $A (\sigma)$,
\begin{equation}
  \psi(\sigma) = s(\sigma) \times A(\sigma),
  \label{eq:phi_1}
\end{equation}
where $\sigma$ is a spin configuration of quantum spin systems. The sign part is fixed. The amplitude is represented by a deep CNN, whose architecture is shown in Fig.~\ref{fig:net}(b), which is denoted as amplitude CNN (aCNN). The input of the aCNN is $\sigma$, which is generated by Markov chain Monte Carlo sampling according to $|\psi(\sigma)|^2$. In the hidden layers of the aCNN, a shortcut connection is used for every two layers, which is packed into a residual block. The concept of a shortcut connection was first proposed in the residual network (ResNet)~\cite{he2015Resnet}, which helps to alleviate the problem of gradient vanishing and exploding encountered in the training process of deep neural networks, and its effectiveness has been widely proved in the research and applications of deep learning. The shortcut connection used in this work is shown in Fig.~\ref{fig:net}(c), where each residual block contains two consecutive convolution operations and the shortcut data are added before the second activation function acts. The pooling layers adopted by traditional CNNs are discarded here because the pooling operations, which resize the shape of the graph, would destroy the translational symmetry of quantum lattice systems.

The connection of the hidden layer in the aCNN is
\begin{equation}
  v_{l}^{c'}=f \left[ \sum_{c=1}^3{C(k_{l-1}^{cc'}, v_{l-1}^c, b^{c'}_{l-1})}\right],
  \label{eq:res_in}
\end{equation}
\begin{equation}
  v_{l+1}^{c'}=f \left[ \sum_{c=1}^3{C(k_{l}^{cc'}, v_{l}^c, b^{c'}_{l}) + v_{l-1}^{c'}} \right],
  \label{eq:res_out}
\end{equation}
in which $v_{l}^c$ is the matrix of layer $l$ with input channel index $c$, $c'$ is the channel index of the output layer of a convolution operation, $v_{l-1}^{c'}$ in Eq.~\eqref{eq:res_out} is the shortcut connection, $f$ is the activation function, ${\theta=\{k, b\}}$ denotes all parameters of the NQS, $k_l$ is convolution kernel, and $b_l$ is the bias of layer $l$. Function $C$ is the standard convolution operation. The two layers constitute a residual block, and the layer number $l =2,4, ...$

The head and tail of the network can be written as
\begin{equation}
  v_{1}^{c} = f \left[ C(k_{0}^{c},v_{0},b^{c}_{0})\right],
  \label{eq:amp_head}
\end{equation}
\begin{equation}
  v_{\mathrm{out}} = f \left[ \sum_{c=1}^3{C(k_{\mathrm{out}-1}, v_{\mathrm{out}-1}^c, b_{\mathrm{out}-1})}\right],
  \label{eq:amp_tail}
\end{equation}
 where $v_{0}$ and $v_{\mathrm{out}}$ are the input and output layers, respectively. The term
\begin{equation}
  a = \mathrm{mean}\left({v}_{\mathrm{out}} \right)
  \label{eq:amp_out}
\end{equation}
 gives the final output of the aCNN, which is the mean value of all outputs of the last convolutional layer. The operation mean is for symmetry, detailed in Sec.~\ref{section:Symmetry}.

Then the wave function can be rewritten as
\begin{equation}
  \psi_{\theta}(\sigma) = s(\sigma) \times \exp \left[ {a_{\theta}(\sigma)} \right].
\end{equation}
Obviously, the number of the network parameters is related to the depth and width of the aCNN and the size of the convolution kernel. In this work, all calculations use the same aCNN architecture with $30$ layers (14 residual blocks) unless otherwise stated. The activation function $f$ is chosen to be ReLU, which maps the data nonlinearly. The middle layer has 3 channels, and the kernel size is $5\times5$. The kernel size does not limit the range of the spin correlation length. As the spin data propagate in the deep CNN, the long-range spin correlation will be gradually built. The actual amount of computation is in proportion to the number of lattice size. The number of (real-valued) parameters of the aCNN is 6538. All tests in this article were performed on an Nvidia 3090 GPU. Actually, when the size of the lattice is large, the Hilbert space of spin configuration is too large to be gone through, which means that the spin configurations involved in the training process are only a very small subset of the Hilbert space. Subsequent tests show the robustness and generalization ability of the aCNN.

\subsection{Optimization}

Theoretically, the approximate ground state of the system can be obtained by optimizing the network's parameters in terms of minimizing the expectation value of energy $E_{\theta}$ of a quantum system. $E_{\theta}$ is used as the loss function of the neural network, and is expressed as
\begin{align}
  E_{\theta} &= \frac{ \langle \psi_{\theta} \left( \sigma \right) \left| H \right| \psi_{\theta} 
  \left( \sigma \right) \rangle}{ \langle \psi_{\theta} \left( \sigma \right) | \psi_{\theta} 
  \left( \sigma \right) \rangle } \nonumber \\
  &=  \langle E_{\mathrm{loc}} \left( \sigma, \theta \right) \rangle_{p(\sigma, \theta)},
  \label{eq:E_theta}
\end{align}
where $E_\mathrm{loc}(\sigma, \theta) = \frac{\langle \sigma|H|\psi_\theta \rangle}{\langle\sigma|\psi_\theta\rangle}$ is the local energy of the system for configuration $\sigma$, $p(\sigma, \theta) = \frac{|\psi_\theta(\sigma)|^2}{\sum_{\sigma}{|\psi_\theta(\sigma)|^2}}$ is the probability distribution, and $\langle \cdot \rangle_{p}$ denotes the expectation with distribution $p$.

The derivative $\nabla_\theta E_\theta$ guides the direction of the optimization. The update of a network's parameters can be expressed as $\delta \theta \propto -\lambda \nabla_\theta E_\theta$, where $\lambda$ is the learning rate. Here, we adopt the idea of the theory of automatic differentiable Monte Carlo~\cite{zhang2023ADVMC} and represent the energy in the following form:
\begin{equation}
  {\mathrm{Es}}
  = {\frac{\langle \frac{\psi_{\theta}^2 \left( \sigma \right)}{\perp(\psi_{\theta}^2 \left( \sigma \right))} \perp({E_\mathrm{loc}(\sigma, \theta)}) 
  \rangle_{\perp(p(\sigma, \theta))}}{\langle \frac{\psi_{\theta}^2 \left( \sigma \right)}{\perp(\psi_{\theta}^2 \left( \sigma \right))}\rangle_{\perp(p(\sigma, \theta))}}},
  \label{eq:Es}
\end{equation}
where $\psi$ denotes $\psi_{\theta} \left( \sigma \right)$. $\perp{\left(x \right)}$ is the detach function, which is defined as $\perp{(x)} = x$ in forward propagation, and $\frac{\partial\perp(x)}{\partial x} = 0$ in backward propagation.
Numerically, the values and first derivative of Eq.~\protect\eqref{eq:E_theta} and Eq.~\protect\eqref{eq:Es} are equal:
\begin{align}
  \perp(\langle E_\mathrm{loc}(\sigma, \theta)\rangle_{p(\sigma, \theta)})
  &= \perp(\mathrm{Es}), \\
  \perp(\nabla_{\theta} {\langle E_\mathrm{loc}(\sigma, \theta)\rangle_{p(\sigma, \theta)}})
  &= \perp(\nabla_{\theta}{\mathrm{Es}}).
\end{align}
The form of expectation Eq.~\protect\eqref{eq:Es} can be encoded into the neural network automatic differentiation framework, which avoids taking the derivative of $p(\sigma, \theta)$ and $E_\mathrm{loc}(\sigma, \theta)$.

In principle, an effective optimization algorithm is the stochastic reconfiguration (SR)~\cite{Sorella2007SR} method. However, the SR requires the inverse of a large matrix of the variational parameters, which limits the number of the parameters and hence the expression ability of the neural network. Here, we only discuss the optimization of amplitude, and choose to use the popular Adam algorithm~\cite{Kingma2015Adam} in deep learning, an algorithm with adaptive learning rate and momentum, which performs well in most tasks.

In the test, the amplitude $a_{\theta} \left( \sigma \right)$ is calculated by the aCNN with real-valued $\theta$, which is initialized randomly by the Kaiming initialization~\cite{he2015KaimingInitilization}, and the sign structure is fixed. The amplitude network is optimized until the loss function no longer drops.

\subsection{Symmetry}
\label{section:Symmetry}

The models simulated in this work are subject to symmetries ${\cal C}_{\rm 4v}$ and translation, unless otherwise stated. Among them, because the scanning of the convolutional kernel on each layer is weight sharing, translation symmetry is an intrinsic property of convolutional connections. Rotation and mirror symmetries can be achieved through data augmentation (${\cal C}_{\rm 4v}$ group operations), as shown in Fig.~\ref{fig:net}(a). After forward propagation, the corresponding results from all symmetric operations are averaged to preserve the symmetries. The symmetrized wave function is
\begin{equation}
  \psi_{\theta}(\sigma)= s(\sigma) \times \exp \left[{\frac{1}{n} 
  \sum_{i=1}^{n}{a(\hat{c}_{\rm 4v}^{(i)}\sigma, \theta)}} \right],
\end{equation}
where $\hat{c}^{(i)}_{\rm 4v}$ denotes a group-element operation of the ${\cal C}_{\rm 4v}$ group, and $n$ is the order of the group.

In addition, the ground states of the Heisenberg model and Heisenberg $J_1$-$J_2$ model both have a conserved quantity $S_z=0$. This can be achieved by hard encoding, which keeps the numbers of up and down spins equal in the sampling process of spin configurations.

Symmetry imposition helps to accelerate the decline of the loss function. In the case of limited computing resources, although ${\cal C}_{\rm 4v}$ symmetry achieved by data augmentation will limit the size of configuration batch allowed in a single training step, data augmentation itself does not limit the number of the network's parameters. Reasonable data augmentation is beneficial to the numerical stability of the optimization process. The accuracy benefits of additional symmetry through data augmentation will be shown in the results below.

\section{Numerical results and discussions}

\subsection{Heisenberg model}

In condensed matter physics, as one of the prototypical quantum many-body systems, the two-dimensional Heisenberg model is a testing ground for various analytical and numerical methods for strongly correlated systems, whose Hamiltonian contains only the nearest-neighbor interactions of the spins, which is unfrustrated. Its ground state is doubtlessly Neel antiferromagnetically ordered. The existing method~\cite{Sandvik2010_VBBasis+LoopUpdate} is very accurate in simulating the ground state of this model, and allows the lattice size to be quite large, providing prefect data for benchmark tests.

The Hamiltonian of the Heisenberg model is
\begin{equation}
  H_{\mathrm{HM}} = \sum_{\langle ij \rangle}{\hat{S_i} \cdot \hat{S_j}},
\end{equation}
where $\hat{S}_i$ is the spin operator on site $i$, and $\langle ij \rangle$ denotes the nearest-neighbor sites $i$ and $j$.

\begin{table}[h!]
  \begin{center}
    \caption{Ground state energies of the Heisenberg model with $L=10,16$ and $24$. Comparison of the results of the aCNN with VMC and QMC; the data are excerpted from Ref.~\protect\cite{Sandvik2010_VBBasis+LoopUpdate}. The relative error is $1-E_{{\rm aCNN}}/E_{\mathrm{QMC}}$.}
    \label{tab:HM}
    \setlength{\tabcolsep}{1.9mm}{
      \begin{tabular}{l r@{.}l  r@{.}l r@{.}l}
        \hline
        \hline
        \multicolumn{1}{l}{Method} & \multicolumn{2}{l}{$L=10$} & \multicolumn{2}{l}{$L=16$} & \multicolumn{2}{l}{$L=24$}  \\
        \hline
        VMC               & \multicolumn{2}{c}{}  & -0&66965      & -0&66922      \\
        QMC                        & -0&67156     & -0&66998      & -0&66961      \\
        aCNN(${\cal C}_{\rm 4v}$)  & -0&671431(2) & -0&669716(1)  & -0&669237(2)  \\
        relative error             &  0&019\%     &  0&039\%      &  0&056\%      \\
        \hline
        \hline
      \end{tabular}
    }
  \end{center}
\end{table}

In this part, we calculate $L\times L$ lattices with $L$ up to 24 under periodic boundary conditions and compare the results with those obtained by VMC and QMC. The sign structure is fixed to the Marshall sign rule~\cite{1955MarshallSign}. As shown in Table~\ref{tab:HM}, the variational optimization results of the aCNN are better than the results of VMC, but there is still a very small gap compared with QMC. The last row of Table~\ref{tab:HM} is the relative error of the energy of the aCNN with respect to that from QMC. The results show that in the direct variational optimization process, the trial wave functions represented by neural networks are better than those of the traditional methods, which may be because of the powerful expression ability of neural networks and the advancement of optimization algorithms.

\begin{figure}[htp!]
  \includegraphics[width=8cm]{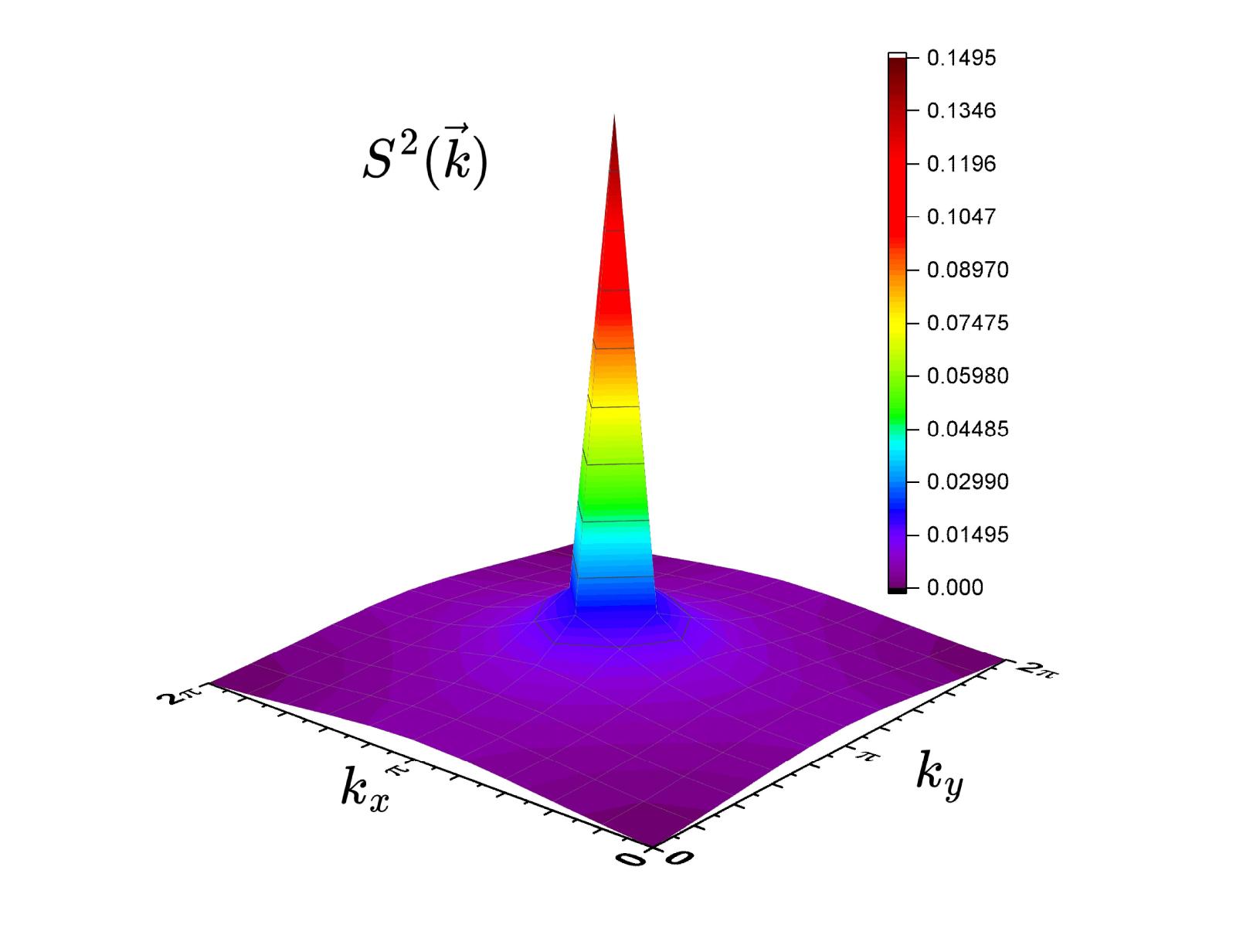}
  \caption{Spin structure factor $S^2(\vec{k})$ as defined in Eq.~\protect\eqref{eq:ssf}. The ${\boldsymbol{\mathrm{k}}}$-space distribution of $S^2(\vec{k})$ of $10\times 10$ square lattice of the Heisenberg model, which peaks at $\vec{k}=(\pi,\pi)$. $k_x$ and $k_y$ are the two components of $\vec{k}$.}
  \label{fig:L10ssf}
\end{figure}

In addition, the spin structure factor
\begin{equation}
  S^2(\vec{k})=\frac{1}{N^2}\sum_{i,j}{\langle \hat{S}_i \cdot \hat{S}_j \rangle e^{i \vec{k}\cdot (\vec{r}_i- \vec{r}_j)}},
  \label{eq:ssf}
\end{equation}
which is related to observable quantities, can be calculated readily, where $\vec{k}$ is a wave vector in $\boldsymbol{\mathrm{k}}$ space. $\vec{r}_i$ is the location of site $i$ in real space and $N$ is the number of lattice sites. As shown in Fig.~\ref{fig:L10ssf}, $S^2(\vec{k})$ is peaked at $\vec{k}=(\pi,\pi)$, indicating the ground state is antiferromagnetically ordered.

\subsection{Transverse field Ising model}

We also test the aCNN for the two-dimensional transverse field Ising model (TFIM) on the square lattice with $L=8$ and $16$. The Hamiltonian is
\begin{equation}
  H_{\mathrm{TFIM}} =-\sum_{\langle i, j \rangle}{\sigma_i^z \sigma_j^z}-\lambda\sum_i{\sigma_i^x},
\end{equation}
where $\sigma^z_i$ and $\sigma^x_i$ are the Pauli operators on site $i$, and $\lambda$ is the transverse field.

All tests are simulations with transverse field $\lambda=3.05$ (close to the critical point $\lambda=3.04438(2)$ ~\cite{DengYoujin2002PRETFIM} of the model), in periodic boundary conditions. The exact sign structure has been employed, namely, positive definite signs. As shown in Table~\ref{tab:TFIM}, the results of the aCNN show excellent agreement with the extrapolated results from the fully augmented tree tensor network (FATTN; $L=8$)~\cite{QinMingpu2022_TreeTensorNetwork-MERA} and fully-augmented matrix product state (FAMPS; $L=16$)~\cite{QinMingpu2023_DMRG-FAMPS}. This test shows that the aCNN works well near the critical point.

\begin{table}[h!]
  \begin{center}
    \caption{Ground state energies of the two-dimensional transverse field Ising model with $L=8$ and $16$. Comparison of the results of the aCNN and the extrapolated results of FATTN~\cite{QinMingpu2022_TreeTensorNetwork-MERA} and FAMPS~\cite{QinMingpu2023_DMRG-FAMPS}.}
    \label{tab:TFIM}
    \setlength{\tabcolsep}{3.2mm}{
      \begin{tabular}{l r@{.}l r@{.}l}
        \hline
        \hline
        \multicolumn{1}{l}{Method} & \multicolumn{2}{l}{$L=8$} & \multicolumn{2}{l}{$L=16$}  \\
        \hline
        aCNN(${\cal C}_{\rm 4v}$) & -3&241642(2)          & -3&239568(2)           \\
        FATTN(extrapolated)       & \multicolumn{2}{c}{}  & -3&23956(2)            \\
        FAMPS(extrapolated)       & -3&24165              & \multicolumn{2}{c}{}   \\
        \hline
        \hline
      \end{tabular}
    }
  \end{center}
\end{table}

\subsection{Heisenberg $J_1$-$J_2$ model}

The above two quantum systems have no frustration, and the sign structures of their exact ground states are known, on which our method has a good performance. Here we try our method on a typical frustrated quantum spin system, the Heisenberg $J_1$-$J_2$ model. The sign structure of its exact ground states is unknown, and even its ground state properties are in debate. Quantum fluctuation and geometric frustration are key factors. Several magnetic long-range orders compete in its most frustrated region, where a quantum spin liquid may emerge.

The Hamiltonian of the Heisenberg $J_1$-$J_2$ model is
\begin{equation}
  H_{\mathrm{J1J2}} = J_1 \sum_{\langle ij \rangle}{\hat{S_i} \cdot \hat{S_j}} + J_2 
  \sum_{\langle \langle ij \rangle \rangle}{\hat{S_i} \cdot \hat{S_j}},
\end{equation}
where $J_1$ and $J_2$ are the coupling strengths of nearest-neighbor sites and next-nearest-neighbor sites (denoted as $\langle \langle ij \rangle \rangle$), respectively. The microscopic details of the model, controlled by the ratio of $J_2 / J_1$, play a decisive role in the type of the ground state and its quantum phase transitions. Here, the model on the square lattice with $L=10$ has been simulated and $J_2 / J_1 = 0.4, 0.45, 0.5$, and $0.55$. The results of the aCNN, the complex-valued CNN~\cite{Choo2019_CNN_ComplexValued}, VMC~\cite{Sorella2013_VMC}, DMRG~\cite{GongShouShu2014_DMRG}, RBM+PP [a method using RBM combined with pair-product (PP) states]~\cite{Nomura2021_RBM+PP} and the deep vision transformer (Deep ViT)~\cite{rende2023DeepViT} are shown in Table~\ref{tab:HMJ1J2}. The sign structure is fixed to the Marshall sign rule, which is not exact when $J_2/J_1 \neq 0$.

\begin{table*}[t!]
  \begin{center}
    \caption{Ground state energies of the Heisenberg $J_1$-$J_2$ model with $L=10$. Comparison of the results of the aCNN with the complex-valued CNN~\cite{Choo2019_CNN_ComplexValued}, VMC~\cite{Sorella2013_VMC}, DMRG~\cite{GongShouShu2014_DMRG}, RBM+PP~\cite{Nomura2021_RBM+PP}, and Deep ViT~\cite{rende2023DeepViT} with $J_2/J_1=0.4, 0.45, 0.5$, and $0.55$. Here, $p$ represents the number of Lanczos steps, and $p = 0$ corresponds to the original variational wave function. VMC($p=\infty$) denotes the variance extrapolated results. In DMRG(4096), `4096' indicates the number of SU(2) states. DMRG($\infty$) is obtained through straight-line energy extrapolation with DMRG truncation error.}
    \label{tab:HMJ1J2}
    \setlength{\tabcolsep}{2mm}{
      \begin{tabular}{l r@{.}l r@{.}l r@{.}l r@{.}l}
        \hline
        \hline
        \multicolumn{1}{l}{Method} & \multicolumn{2}{c}{$J_2/J_1=0.4$} & \multicolumn{2}{c}{$J_2/J_1=0.45$} & \multicolumn{2}{c}{$J_2/J_1=0.5$} & \multicolumn{2}{c}{$J_2/J_1=0.55$} \\
        \hline
        VMC($p=0$)                   & -0&52188(1)   & -0&50811(1)   & -0&49521(1)   & -0&48335(1)   \\
        VMC($p=1$)                   & -0&52368(1)   & -0&50973(1)   & -0&49718(1)   & -0&48622(1)   \\
        VMC($p=2$)                   & -0&5240(1)    & -0&51001(1)   & -0&49755(1)   & -0&48693(3)   \\
        VMC($p=\infty$)             & -0&52429(2)   & -0&51017(2)   & -0&49781(2)   & -0&48766(6)   \\
        DMRG(4096)                & -0&521487     & -0&507193     & -0&495044     & -0&484890     \\
        DMRG(6144)                & -0&522043     & -0&507677     & -0&495301     & -0&485239     \\
        DMRG(8192)                & -0&522391     & -0&507976     & -0&495530     & -0&485434     \\
        DMRG($\infty$)            & -0&5253       & -0&5110       & -0&4988       & -0&4880       \\
        RBM+PP                    & \multicolumn{2}{c}{} & \multicolumn{2}{c}{} & -0&497629(1) & \multicolumn{2}{c}{} \\
        Deep ViT                  & \multicolumn{2}{c}{} & \multicolumn{2}{c}{} & -0&497634(1) & \multicolumn{2}{c}{} \\
        complex-valued CNN        & -0&52371(1)   & -0&50905(1)   & -0&49516(1)   & -0&48277(1)   \\
        aCNN(${\cal C}_{\rm 4v}$) & -0&523999(4)  & -0&509207(5)  & -0&495627(6)  & -0&483490(5)  \\
        aCNN(${\cal C}_{4}$)      & -0&523872(4)  & \multicolumn{2}{c}{} & \multicolumn{2}{c}{} & -0&483173(5) \\
        \hline
        \hline
      \end{tabular}
    }
  \end{center}
\end{table*}

As shown in Table~\ref{tab:HMJ1J2}, the energy obtained by the aCNN is better than that obtained by VMC($p=0$) and especially by the complex-valued CNN. This shows that the wave function of the aCNN has advantages, and in direct variational optimization, NQS optimized by the deep learning algorithm is a better form than the traditional trial wave function used by VMC. Compared with the complex-valued CNN, the results show that it is difficult to optimize both sign and amplitude for the complex-valued CNN.

Since the mirror reflection symmetry is not used in the simulation with the complex-valued CNN, the test using ${\cal C}_{4}$ symmetry only is also done in this work to eliminate the influence of different symmetries. As an example, the Heisenberg $J_1$-$J_2$ model ($L=10$) with $J_2=0.4$ and $0.55$ has been simulated. The energy obtained in this work is significantly better than that of the complex-valued CNN, as shown in Table~\ref{tab:HMJ1J2}. The complex-valued CNN uses SR to optimize the network, which in principle can optimize the sign structure and amplitude of the wave function simultaneously. The comparison results show that it seems that the sign structure is not well optimized in the complex-valued CNN since its energy is even worse than that from the aCNN with an incorrect sign structure.

Comparing the network architectures of the aCNN and the complex-valued CNN, the latter network is wide and shallow while the former is narrow and deep relatively, and the former has residual connections additionally. The different architectures have a certain impact on their expression ability. However, it is important to note that the numbers of their parameters are almost equal. The complex-valued CNN contains 3838 complex-valued parameters, while the aCNN contains 6538 real-valued parameters.

The results of VMC and DMRG are listed in Table~\ref{tab:HMJ1J2}. The energies of the aCNN are even better than those of VMC($p=1$) (with $p=1$ Lanczos correction), and comparable to that of VMC($p=2$) when $J_2/J_1=0.4$. As the frustration increases, the advantage of the aCNN gradually decreases. When the frustration is the strongest at $J_2/J_1=0.55$, the performance of the aCNN deteriorates. This implies that the Marshall sign rule is significantly broken by frustration here. The ground-state behavior of the system is most controversial near the strongest frustration point $J_2/J_1=0.55$. Because the combined effect of frustration and quantum fluctuation destroys the long-range antiferromagnetic order, the ground state behaves as nonmagnetic. However, the nature of this quantum phase remains controversial. The above results illustrate the importance of sign structure optimization, which will be studied in the future.

In addition, comparing the results of the aCNN(${\cal C}_{\rm 4v}$) with that of the aCNN(${\cal C}_{4}$) in Table~\ref{tab:HMJ1J2}, it is clear whether the use of mirror reflection symmetry is important for the simulation. This shows again that symmetry imposition helps.

\subsection{Applicability of sign structures and sources of error}

In the calculations of the Heisenberg $J_1$-$J_2$ model discussed earlier, the sign structure is fixed to the Marshall sign rule (checkerboard-patterned), which is not exact when $J_2 \neq 0$. It is essential to understand the applicability of sign structures and to figure out the sources of error in the aCNN. 

\begin{figure}[b!]
  \includegraphics[width=8.6cm]{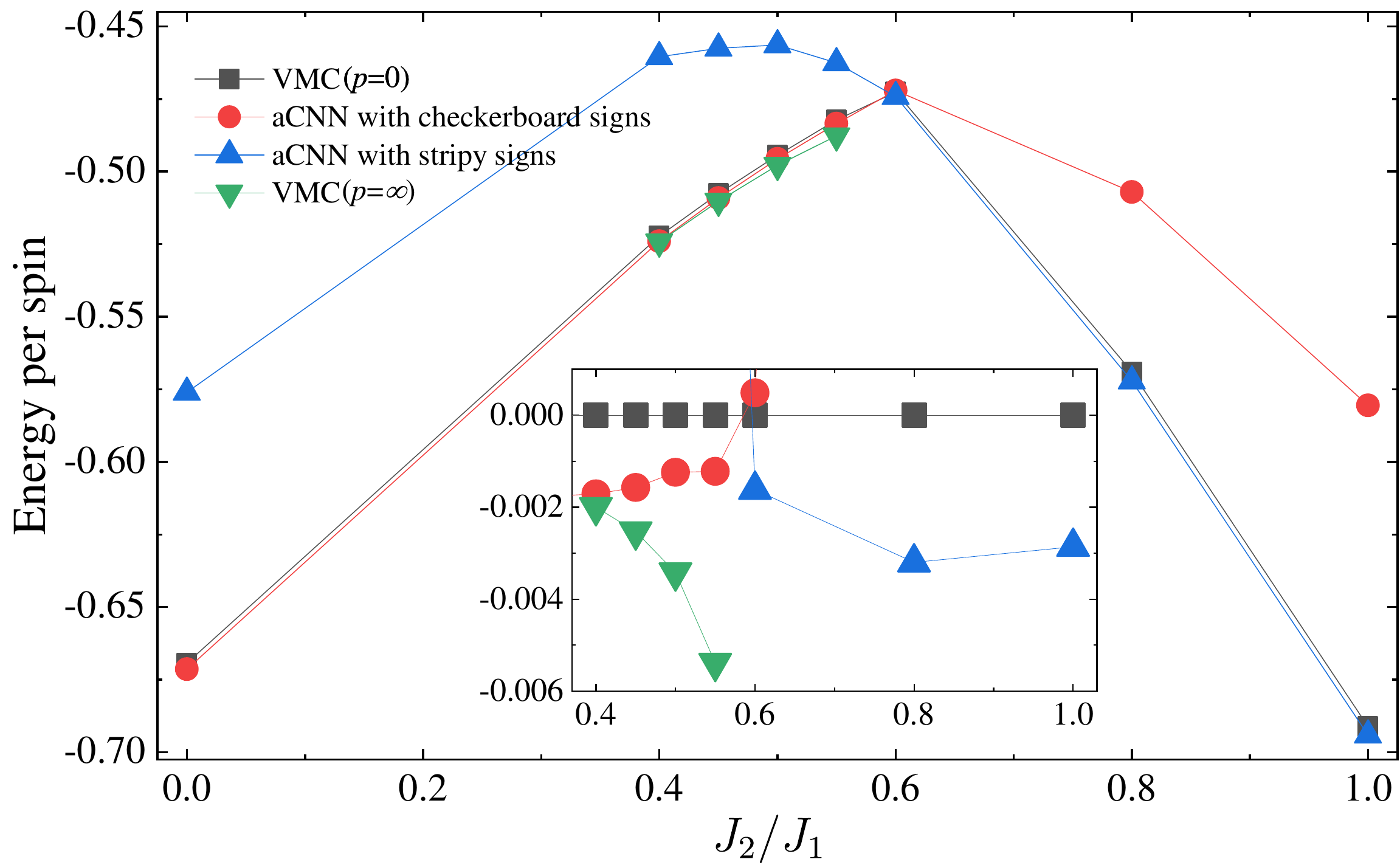}
  \caption{Ground state energies of the Heisenberg $J_1$-$J_2$ model with $L=10$. The aCNN is optimized with different Marshall sign structures, the checkerboard-patterned (exact for $J_2=0$) and the stripe-patterned (exact for $J_1=0$). The VMC($p=0$) results are from Ref.~\cite{Choo2019_CNN_ComplexValued}. VMC($p=\infty$) denotes the variance extrapolated results from Ref.~\cite{Sorella2013_VMC}, which can be regarded as the exact ground state energies. The inset shows $E - E(\mathrm{VMC}(p=0))$, the relative energies with respect to the VMC($p=0$) energies, for an enlarged view.}
  \label{fig:L10_nH_sH_Jp0_1}
\end{figure}

Here, the Heisenberg $J_1$-$J_2$ model ($L=10$) is simulated with different Marshall sign structures, the checkerboard-patterned (exact for $J_2=0$) and the stripe-patterned (exact for $J_1=0$). The results are depicted in Fig.~\ref{fig:L10_nH_sH_Jp0_1}, together with the results of VMC($p=0$) (provided by F. Ferrari and F. Becca in Ref.~\cite{Choo2019_CNN_ComplexValued}) and VMC($p=\infty$)~\cite{Sorella2013_VMC}. VMC($p=\infty$) denotes the variance extrapolated results, which can be regarded as the exact ground state energies.

As depicted in Fig.~\ref{fig:L10_nH_sH_Jp0_1}, the ground state energies obtained from the aCNN consistently outperform those from VMC($p=0$), and are close to the results of VMC($p=\infty$) at $J_2/J_1=0.4, 0.45, 0.5$, and $0.55$, which indicates that even though the frustration is strong ($J_2/J_1=0.5$ and $0.55$), the sign structures fixed according to the Marshall sign rule remain highly close to the exact ground state sign structures. This is consistent with the estimation from Refs.~\cite{Claudio2020_NQS_SignProblem,Ivanov1997_MSR_exact_in_J1J2}. In addition, when $J_2/J_1 < 0.6$, the energies obtained by the aCNN with the stripy signs are significantly higher than those from VMC($p=0$), and when $J_2/J_1 > 0.6$, the energies obtained by the aCNN with the checkerboard signs are significantly higher than those from VMC($p=0$). This demonstrates that when the sign structure employed by the aCNN diverges significantly from that of the exact ground state, the errors in the results will become very large, underscoring the importance of a proper sign structure in obtaining accurate results.

To pinpoint the actual source of error in the aCNN, we conduct simulations of the Heisenberg $J_1$-$J_2$ model with $J_2/J_1=0.5$ on a $6\times6$ square lattice, and the Heisenberg model on a triangular lattice (HMTL) of $36$ sites, whose exact ground states are available through exact diagonalization (ED) methods. Here, the aCNN is optimized with the fixed sign structures of the exact ground states.

These two models are both strongly frustrated. In particular, the geometric frustration inherent in the triangular lattice makes the amplitude more complex. This is a challenge for the expressive power of neural networks. To improve the accuracy of the result, a deep neural network with ${\cal C}_{\rm 6v}$ symmetry, consisting of 100 layers, is employed in the simulation for the HMTL.

\begin{table}[h!]
  \begin{center}
    \caption{Ground state energies of the Heisenberg $J_1$-$J_2$ model with $J_2/J_1=0.5$ on a $6\times6$ square lattice and the Heisenberg model on a triangular lattice (HMTL) of 36 sites. The aCNN is optimized with the exact sign structure from ED. Results of LCN are from Ref.~\cite{fu2022latticeCNN}, and GNN/GNN-2 are from Ref.~\cite{kochkov2021GNN}. The column P/M indicates the number of parameters (P) in millions (M). The relative error is $1-E_{\mathrm{aCNN}}/E_{\mathrm{ED}}$.}
    \label{tab:another_frustration}
    \setlength{\tabcolsep}{1.2mm}{
      \begin{tabular}{l c|c r@{.}l r@{.}l c|c r@{.}l r@{.}l}
        \hline
        \hline
        \multicolumn{1}{l}{Method} & & & \multicolumn{2}{l}{$J_1$-$J_2$}  & \multicolumn{2}{l}{P/M} & & & \multicolumn{2}{l}{HMTL}  & \multicolumn{2}{l}{P/M} \\
        \hline
        ED          & & & -0&503810     & \multicolumn{2}{c}{ }  & & & -0&560374        & \multicolumn{2}{c}{ }  \\
        GNN         & & & -0&5022(4)    & 0&86                   & & & -0&55892(2)      & 0&86    \\
        GNN-2       & & & -0&5023(5)    & \multicolumn{2}{c}{ }  & & & -0&55949(2)      & \multicolumn{2}{c}{ }   \\
        LCN         & & & -0&5022(2)    & 0&28                   & & & -0&5601(4)       & 0&29  \\
        aCNN        & & & -0&503258(4)  & 0&0065                 & & & -0&559214(6)     & 0&022   \\
        rel. error & & & 0&110\% & \multicolumn{2}{c}{ }  & & & 0&207\%         & \multicolumn{2}{c}{ }  \\
        \hline
        \hline
      \end{tabular}
    }
  \end{center}
\end{table}

The results of the aCNN are shown in Table~\ref{tab:another_frustration}, together with the results of the lattice convolutional network (LCN)~\cite{fu2022latticeCNN} and graph neural network (GNN)~\cite{kochkov2021GNN} for comparison. As evident, the aCNN achieves precision levels comparable to those of GNN, GNN-2, and LCN, yet with significantly lesser parameters than each of them. Comparing the relative errors listed in Table~\ref{tab:HM} and Table~\ref{tab:another_frustration}, the relative errors for frustrated systems are higher than those for unfrustrated systems. This implies that the presence of strong frustration would result in heightened complexity of the amplitude of the ground state wave function, and deeper neural networks are necessary for such systems.

In this paper, the aCNN is optimized with fixed sign structure. Sometimes there are known sign structures, such as the Heisenberg model and the transverse field Ising model. At other times, approximate ground state wave functions and their sign structures can be obtained through alternative methods, such as mean-field theories, low-parameter VMC~\cite{Sorella2013_VMC}, or tensor network states~\cite{Zheng-ChengGu2022_TensorNetwork+PEPS}. In such cases, using the aCNN could achieve more accurate ground state wave functions.

Furthermore, the ground state wave functions of the models calculated in this paper are all real-valued. When the Hamiltonian contains complex numbers, or when the wave function evolves in real time, the system's wave function will be complex-valued with continuous phases. In those cases, the aCNN needs to be used with a phase network, which is worth further research in the future.

\section{Summary and outlook}

In summary, we have proposed a quantum many-body wave function, named aCNN, which is a multiplication of a real-valued amplitude neural network (a deep CNN with residual blocks) and a fixed sign structure. The ground state energy has been obtained by variational optimization of the aCNN with the Adam algorithm. Our method is tested on typical quantum many-body systems. The obtained ground state energies are better than or comparable to those from traditional VMC methods, DMRG, and the complex-valued CNN.

The results provide specific evidence to support the viewpoint that the deep aCNN can be a competitive wave function ansatz for describing the amplitude of the ground state wave functions and properties of quantum many-body systems compared with traditional VMC methods. Importantly, this work shows that the aCNN can obtain better energies than the complex-valued CNN with the same number of variational parameters, which implies that it is not easy to optimize both the sign structure and amplitude simultaneously. And it supports the viewpoint that it is worth trying to use different tools to represent the sign structure and amplitude, respectively, suggested also by Ref.~\cite{westerhout2023SignStructure}. Here we show that CNN is a good choice for the amplitude part.

In the long run, it is worth continuing to explore the use of more advanced artificial intelligence technologies to optimize the ground state wave function of quantum many-body systems, especially quantum frustrated systems and strongly correlated fermionic systems with a sign problem. Particularly, it is worthwhile to explore a way to optimizing the sign structure with NQSs in the future. 

The main code and data for the aCNN are available \footnote{The main code and data for the aCNN are available on GitHub at https://github.com/rqHe1/aCNN.}.

\begin{acknowledgments}
This work was supported by the National Natural Science Foundation of China (Grant No. 11934020). Computational resources were provided by the Physical Laboratory of High Performance Computing at Renmin University of China.
\end{acknowledgments}

\bibliography{ampcnn}

\begin{thebibliography}{56}%
\makeatletter
\providecommand \@ifxundefined [1]{%
 \@ifx{#1\undefined}
}%
\providecommand \@ifnum [1]{%
 \ifnum #1\expandafter \@firstoftwo
 \else \expandafter \@secondoftwo
 \fi
}%
\providecommand \@ifx [1]{%
 \ifx #1\expandafter \@firstoftwo
 \else \expandafter \@secondoftwo
 \fi
}%
\providecommand \natexlab [1]{#1}%
\providecommand \enquote  [1]{``#1''}%
\providecommand \bibnamefont  [1]{#1}%
\providecommand \bibfnamefont [1]{#1}%
\providecommand \citenamefont [1]{#1}%
\providecommand \href@noop [0]{\@secondoftwo}%
\providecommand \href [0]{\begingroup \@sanitize@url \@href}%
\providecommand \@href[1]{\@@startlink{#1}\@@href}%
\providecommand \@@href[1]{\endgroup#1\@@endlink}%
\providecommand \@sanitize@url [0]{\catcode `\\12\catcode `\$12\catcode
  `\&12\catcode `\#12\catcode `\^12\catcode `\_12\catcode `\%12\relax}%
\providecommand \@@startlink[1]{}%
\providecommand \@@endlink[0]{}%
\providecommand \url  [0]{\begingroup\@sanitize@url \@url }%
\providecommand \@url [1]{\endgroup\@href {#1}{\urlprefix }}%
\providecommand \urlprefix  [0]{URL }%
\providecommand \Eprint [0]{\href }%
\providecommand \doibase [0]{https://doi.org/}%
\providecommand \selectlanguage [0]{\@gobble}%
\providecommand \bibinfo  [0]{\@secondoftwo}%
\providecommand \bibfield  [0]{\@secondoftwo}%
\providecommand \translation [1]{[#1]}%
\providecommand \BibitemOpen [0]{}%
\providecommand \bibitemStop [0]{}%
\providecommand \bibitemNoStop [0]{.\EOS\space}%
\providecommand \EOS [0]{\spacefactor3000\relax}%
\providecommand \BibitemShut  [1]{\csname bibitem#1\endcsname}%
\let\auto@bib@innerbib\@empty
\bibitem [{\citenamefont {Lam}\ \emph {et~al.}(2023)\citenamefont {Lam},
  \citenamefont {Sanchez-Gonzalez}, \citenamefont {Willson}, \citenamefont
  {Wirnsberger}, \citenamefont {Fortunato}, \citenamefont {Alet}, \citenamefont
  {Ravuri}, \citenamefont {Ewalds}, \citenamefont {Eaton-Rosen}, \citenamefont
  {Hu}, \citenamefont {Merose}, \citenamefont {Hoyer}, \citenamefont {Holland},
  \citenamefont {Vinyals}, \citenamefont {Stott}, \citenamefont {Pritzel},
  \citenamefont {Mohamed},\ and\ \citenamefont
  {Battaglia}}]{lam2023WeatherForecasting}%
  \BibitemOpen
  \bibfield  {author} {\bibinfo {author} {\bibfnamefont {R.}~\bibnamefont
  {Lam}}, \bibinfo {author} {\bibfnamefont {A.}~\bibnamefont
  {Sanchez-Gonzalez}}, \bibinfo {author} {\bibfnamefont {M.}~\bibnamefont
  {Willson}}, \bibinfo {author} {\bibfnamefont {P.}~\bibnamefont
  {Wirnsberger}}, \bibinfo {author} {\bibfnamefont {M.}~\bibnamefont
  {Fortunato}}, \bibinfo {author} {\bibfnamefont {F.}~\bibnamefont {Alet}},
  \bibinfo {author} {\bibfnamefont {S.}~\bibnamefont {Ravuri}}, \bibinfo
  {author} {\bibfnamefont {T.}~\bibnamefont {Ewalds}}, \bibinfo {author}
  {\bibfnamefont {Z.}~\bibnamefont {Eaton-Rosen}}, \bibinfo {author}
  {\bibfnamefont {W.}~\bibnamefont {Hu}}, \bibinfo {author} {\bibfnamefont
  {A.}~\bibnamefont {Merose}}, \bibinfo {author} {\bibfnamefont
  {S.}~\bibnamefont {Hoyer}}, \bibinfo {author} {\bibfnamefont
  {G.}~\bibnamefont {Holland}}, \bibinfo {author} {\bibfnamefont
  {O.}~\bibnamefont {Vinyals}}, \bibinfo {author} {\bibfnamefont
  {J.}~\bibnamefont {Stott}}, \bibinfo {author} {\bibfnamefont
  {A.}~\bibnamefont {Pritzel}}, \bibinfo {author} {\bibfnamefont
  {S.}~\bibnamefont {Mohamed}},\ and\ \bibinfo {author} {\bibfnamefont
  {P.}~\bibnamefont {Battaglia}},\ }\bibfield  {title} {\bibinfo {title}
  {Learning skillful medium-range global weather forecasting},\ }\href
  {https://doi.org/10.1126/science.adi2336} {\bibfield  {journal} {\bibinfo
  {journal} {Science}\ }\textbf {\bibinfo {volume} {382}},\ \bibinfo {pages}
  {1416} (\bibinfo {year} {2023})}\BibitemShut {NoStop}%
\bibitem [{\citenamefont {Jumper}\ \emph {et~al.}(2021)\citenamefont {Jumper},
  \citenamefont {Evans}, \citenamefont {Pritzel}, \citenamefont {Green},
  \citenamefont {Figurnov}, \citenamefont {Ronneberger}, \citenamefont
  {Tunyasuvunakool}, \citenamefont {Bates}, \citenamefont {Žídek},
  \citenamefont {Potapenko}, \citenamefont {Bridgland}, \citenamefont {Meyer},
  \citenamefont {Kohl}, \citenamefont {Ballard}, \citenamefont {Cowie},
  \citenamefont {Romera-Paredes}, \citenamefont {Nikolov}, \citenamefont
  {Jain}, \citenamefont {Adler}, \citenamefont {Back}, \citenamefont
  {Petersen}, \citenamefont {Reiman}, \citenamefont {Clancy}, \citenamefont
  {Zielinski}, \citenamefont {Steinegger}, \citenamefont {Pacholska},
  \citenamefont {Berghammer}, \citenamefont {Bodenstein}, \citenamefont
  {Silver}, \citenamefont {Vinyals}, \citenamefont {Senior}, \citenamefont
  {Kavukcuoglu}, \citenamefont {Kohli},\ and\ \citenamefont
  {Hassabis}}]{2021AlphaFold}%
  \BibitemOpen
  \bibfield  {author} {\bibinfo {author} {\bibfnamefont {J.}~\bibnamefont
  {Jumper}}, \bibinfo {author} {\bibfnamefont {R.}~\bibnamefont {Evans}},
  \bibinfo {author} {\bibfnamefont {A.}~\bibnamefont {Pritzel}}, \bibinfo
  {author} {\bibfnamefont {T.}~\bibnamefont {Green}}, \bibinfo {author}
  {\bibfnamefont {M.}~\bibnamefont {Figurnov}}, \bibinfo {author}
  {\bibfnamefont {O.}~\bibnamefont {Ronneberger}}, \bibinfo {author}
  {\bibfnamefont {K.}~\bibnamefont {Tunyasuvunakool}}, \bibinfo {author}
  {\bibfnamefont {R.}~\bibnamefont {Bates}}, \bibinfo {author} {\bibfnamefont
  {A.}~\bibnamefont {Žídek}}, \bibinfo {author} {\bibfnamefont
  {A.}~\bibnamefont {Potapenko}}, \bibinfo {author} {\bibfnamefont
  {A.}~\bibnamefont {Bridgland}}, \bibinfo {author} {\bibfnamefont
  {C.}~\bibnamefont {Meyer}}, \bibinfo {author} {\bibfnamefont {S.~A.~A.}\
  \bibnamefont {Kohl}}, \bibinfo {author} {\bibfnamefont {A.~J.}\ \bibnamefont
  {Ballard}}, \bibinfo {author} {\bibfnamefont {A.}~\bibnamefont {Cowie}},
  \bibinfo {author} {\bibfnamefont {B.}~\bibnamefont {Romera-Paredes}},
  \bibinfo {author} {\bibfnamefont {S.}~\bibnamefont {Nikolov}}, \bibinfo
  {author} {\bibfnamefont {R.}~\bibnamefont {Jain}}, \bibinfo {author}
  {\bibfnamefont {J.}~\bibnamefont {Adler}}, \bibinfo {author} {\bibfnamefont
  {T.}~\bibnamefont {Back}}, \bibinfo {author} {\bibfnamefont {S.}~\bibnamefont
  {Petersen}}, \bibinfo {author} {\bibfnamefont {D.}~\bibnamefont {Reiman}},
  \bibinfo {author} {\bibfnamefont {E.}~\bibnamefont {Clancy}}, \bibinfo
  {author} {\bibfnamefont {M.}~\bibnamefont {Zielinski}}, \bibinfo {author}
  {\bibfnamefont {M.}~\bibnamefont {Steinegger}}, \bibinfo {author}
  {\bibfnamefont {M.}~\bibnamefont {Pacholska}}, \bibinfo {author}
  {\bibfnamefont {T.}~\bibnamefont {Berghammer}}, \bibinfo {author}
  {\bibfnamefont {S.}~\bibnamefont {Bodenstein}}, \bibinfo {author}
  {\bibfnamefont {D.}~\bibnamefont {Silver}}, \bibinfo {author} {\bibfnamefont
  {O.}~\bibnamefont {Vinyals}}, \bibinfo {author} {\bibfnamefont {A.~W.}\
  \bibnamefont {Senior}}, \bibinfo {author} {\bibfnamefont {K.}~\bibnamefont
  {Kavukcuoglu}}, \bibinfo {author} {\bibfnamefont {P.}~\bibnamefont {Kohli}},\
  and\ \bibinfo {author} {\bibfnamefont {D.}~\bibnamefont {Hassabis}},\
  }\bibfield  {title} {\bibinfo {title} {Highly accurate protein structure
  prediction with alphafold},\ }\href
  {https://doi.org/10.1038/s41586-021-03819-2} {\bibfield  {journal} {\bibinfo
  {journal} {Nature}\ }\textbf {\bibinfo {volume} {596}},\ \bibinfo {pages}
  {583} (\bibinfo {year} {2021})}\BibitemShut {NoStop}%
\bibitem [{\citenamefont {Press}(2017)}]{2017Introduction}%
  \BibitemOpen
  \bibfield  {author} {\bibinfo {author} {\bibfnamefont {C.~U.}\ \bibnamefont
  {Press}},\ }\bibfield  {title} {\bibinfo {title} {Introduction to many-body
  physics},\ }\href@noop {} {\bibfield  {journal} {\bibinfo  {journal} {Physics
  Today}\ }\textbf {\bibinfo {volume} {70}},\ \bibinfo {pages} {59} (\bibinfo
  {year} {2017})}\BibitemShut {NoStop}%
\bibitem [{\citenamefont {Huang}\ \emph {et~al.}(2022)\citenamefont {Huang},
  \citenamefont {Kueng}, \citenamefont {Torlai}, \citenamefont {Albert},\ and\
  \citenamefont {Preskill}}]{Huang2022QuantumCalc}%
  \BibitemOpen
  \bibfield  {author} {\bibinfo {author} {\bibfnamefont {H.-Y.}\ \bibnamefont
  {Huang}}, \bibinfo {author} {\bibfnamefont {R.}~\bibnamefont {Kueng}},
  \bibinfo {author} {\bibfnamefont {G.}~\bibnamefont {Torlai}}, \bibinfo
  {author} {\bibfnamefont {V.~V.}\ \bibnamefont {Albert}},\ and\ \bibinfo
  {author} {\bibfnamefont {J.}~\bibnamefont {Preskill}},\ }\bibfield  {title}
  {\bibinfo {title} {Provably efficient machine learning for quantum many-body
  problems},\ }\href {https://doi.org/10.1126/science.abk3333} {\bibfield
  {journal} {\bibinfo  {journal} {Science}\ }\textbf {\bibinfo {volume}
  {377}},\ \bibinfo {pages} {eabk3333} (\bibinfo {year} {2022})}\BibitemShut
  {NoStop}%
\bibitem [{\citenamefont {Carleo}\ \emph {et~al.}(2019)\citenamefont {Carleo},
  \citenamefont {Cirac}, \citenamefont {Cranmer}, \citenamefont {Daudet},
  \citenamefont {Schuld}, \citenamefont {Tishby}, \citenamefont
  {Vogt-Maranto},\ and\ \citenamefont
  {Zdeborov\'a}}]{Carleo2019Review_AI_Physics}%
  \BibitemOpen
  \bibfield  {author} {\bibinfo {author} {\bibfnamefont {G.}~\bibnamefont
  {Carleo}}, \bibinfo {author} {\bibfnamefont {I.}~\bibnamefont {Cirac}},
  \bibinfo {author} {\bibfnamefont {K.}~\bibnamefont {Cranmer}}, \bibinfo
  {author} {\bibfnamefont {L.}~\bibnamefont {Daudet}}, \bibinfo {author}
  {\bibfnamefont {M.}~\bibnamefont {Schuld}}, \bibinfo {author} {\bibfnamefont
  {N.}~\bibnamefont {Tishby}}, \bibinfo {author} {\bibfnamefont
  {L.}~\bibnamefont {Vogt-Maranto}},\ and\ \bibinfo {author} {\bibfnamefont
  {L.}~\bibnamefont {Zdeborov\'a}},\ }\bibfield  {title} {\bibinfo {title}
  {Machine learning and the physical sciences},\ }\href
  {https://doi.org/10.1103/RevModPhys.91.045002} {\bibfield  {journal}
  {\bibinfo  {journal} {Rev. Mod. Phys.}\ }\textbf {\bibinfo {volume} {91}},\
  \bibinfo {pages} {045002} (\bibinfo {year} {2019})}\BibitemShut {NoStop}%
\bibitem [{\citenamefont {Carleo}\ and\ \citenamefont
  {Troyer}(2017)}]{GiuseppeCarleo2017_RBM}%
  \BibitemOpen
  \bibfield  {author} {\bibinfo {author} {\bibfnamefont {G.}~\bibnamefont
  {Carleo}}\ and\ \bibinfo {author} {\bibfnamefont {M.}~\bibnamefont
  {Troyer}},\ }\bibfield  {title} {\bibinfo {title} {Solving the quantum
  many-body problem with artificial neural networks},\ }\href
  {https://doi.org/10.1126/science.aag2302} {\bibfield  {journal} {\bibinfo
  {journal} {Science}\ }\textbf {\bibinfo {volume} {355}},\ \bibinfo {pages}
  {602} (\bibinfo {year} {2017})}\BibitemShut {NoStop}%
\bibitem [{\citenamefont {Nomura}(2022)}]{Nomura2022ParametersRBM_NNQS}%
  \BibitemOpen
  \bibfield  {author} {\bibinfo {author} {\bibfnamefont {Y.}~\bibnamefont
  {Nomura}},\ }\bibfield  {title} {\bibinfo {title} {Investigating network
  parameters in neural-network quantum states},\ }\href
  {https://doi.org/10.7566/JPSJ.91.054709} {\bibfield  {journal} {\bibinfo
  {journal} {Journal of the Physical Society of Japan}\ }\textbf {\bibinfo
  {volume} {91}},\ \bibinfo {pages} {054709} (\bibinfo {year}
  {2022})}\BibitemShut {NoStop}%
\bibitem [{\citenamefont {Zhang}\ \emph
  {et~al.}(2023{\natexlab{a}})\citenamefont {Zhang}, \citenamefont {Xu},
  \citenamefont {Wu}, \citenamefont {Balachandran},\ and\ \citenamefont
  {Poletti}}]{Zhang2023_LocalSequentialUpdateNNQS}%
  \BibitemOpen
  \bibfield  {author} {\bibinfo {author} {\bibfnamefont {W.}~\bibnamefont
  {Zhang}}, \bibinfo {author} {\bibfnamefont {X.}~\bibnamefont {Xu}}, \bibinfo
  {author} {\bibfnamefont {Z.}~\bibnamefont {Wu}}, \bibinfo {author}
  {\bibfnamefont {V.}~\bibnamefont {Balachandran}},\ and\ \bibinfo {author}
  {\bibfnamefont {D.}~\bibnamefont {Poletti}},\ }\bibfield  {title} {\bibinfo
  {title} {Ground state search by local and sequential updates of neural
  network quantum states},\ }\href
  {https://doi.org/10.1103/PhysRevB.107.165149} {\bibfield  {journal} {\bibinfo
   {journal} {Phys. Rev. B}\ }\textbf {\bibinfo {volume} {107}},\ \bibinfo
  {pages} {165149} (\bibinfo {year} {2023}{\natexlab{a}})}\BibitemShut
  {NoStop}%
\bibitem [{\citenamefont {Roth}\ \emph {et~al.}(2023)\citenamefont {Roth},
  \citenamefont {Szab\'o},\ and\ \citenamefont {MacDonald}}]{roth2023_GCNN}%
  \BibitemOpen
  \bibfield  {author} {\bibinfo {author} {\bibfnamefont {C.}~\bibnamefont
  {Roth}}, \bibinfo {author} {\bibfnamefont {A.}~\bibnamefont {Szab\'o}},\ and\
  \bibinfo {author} {\bibfnamefont {A.~H.}\ \bibnamefont {MacDonald}},\
  }\bibfield  {title} {\bibinfo {title} {High-accuracy variational monte carlo
  for frustrated magnets with deep neural networks},\ }\href
  {https://doi.org/10.1103/PhysRevB.108.054410} {\bibfield  {journal} {\bibinfo
   {journal} {Phys. Rev. B}\ }\textbf {\bibinfo {volume} {108}},\ \bibinfo
  {pages} {054410} (\bibinfo {year} {2023})}\BibitemShut {NoStop}%
\bibitem [{\citenamefont {Viteritti}\ \emph {et~al.}(2023)\citenamefont
  {Viteritti}, \citenamefont {Rende},\ and\ \citenamefont
  {Becca}}]{Luciano2023_Transformer_1D_J1J2}%
  \BibitemOpen
  \bibfield  {author} {\bibinfo {author} {\bibfnamefont {L.~L.}\ \bibnamefont
  {Viteritti}}, \bibinfo {author} {\bibfnamefont {R.}~\bibnamefont {Rende}},\
  and\ \bibinfo {author} {\bibfnamefont {F.}~\bibnamefont {Becca}},\ }\bibfield
   {title} {\bibinfo {title} {Transformer variational wave functions for
  frustrated quantum spin systems},\ }\href
  {https://doi.org/10.1103/PhysRevLett.130.236401} {\bibfield  {journal}
  {\bibinfo  {journal} {Phys. Rev. Lett.}\ }\textbf {\bibinfo {volume} {130}},\
  \bibinfo {pages} {236401} (\bibinfo {year} {2023})}\BibitemShut {NoStop}%
\bibitem [{\citenamefont {Zhang}\ and\ \citenamefont
  {Di~Ventra}(2023)}]{Zhang2023Transformer_NQS_TFIM}%
  \BibitemOpen
  \bibfield  {author} {\bibinfo {author} {\bibfnamefont {Y.-H.}\ \bibnamefont
  {Zhang}}\ and\ \bibinfo {author} {\bibfnamefont {M.}~\bibnamefont
  {Di~Ventra}},\ }\bibfield  {title} {\bibinfo {title} {Transformer quantum
  state: A multipurpose model for quantum many-body problems},\ }\href
  {https://doi.org/10.1103/PhysRevB.107.075147} {\bibfield  {journal} {\bibinfo
   {journal} {Phys. Rev. B}\ }\textbf {\bibinfo {volume} {107}},\ \bibinfo
  {pages} {075147} (\bibinfo {year} {2023})}\BibitemShut {NoStop}%
\bibitem [{\citenamefont {Reh}\ \emph {et~al.}(2023)\citenamefont {Reh},
  \citenamefont {Schmitt},\ and\ \citenamefont
  {G\"arttner}}]{reh2023DetailsSymmetrization}%
  \BibitemOpen
  \bibfield  {author} {\bibinfo {author} {\bibfnamefont {M.}~\bibnamefont
  {Reh}}, \bibinfo {author} {\bibfnamefont {M.}~\bibnamefont {Schmitt}},\ and\
  \bibinfo {author} {\bibfnamefont {M.}~\bibnamefont {G\"arttner}},\ }\bibfield
   {title} {\bibinfo {title} {Optimizing design choices for neural quantum
  states},\ }\href {https://doi.org/10.1103/PhysRevB.107.195115} {\bibfield
  {journal} {\bibinfo  {journal} {Phys. Rev. B}\ }\textbf {\bibinfo {volume}
  {107}},\ \bibinfo {pages} {195115} (\bibinfo {year} {2023})}\BibitemShut
  {NoStop}%
\bibitem [{\citenamefont {Liang}\ \emph {et~al.}(2018)\citenamefont {Liang},
  \citenamefont {Liu}, \citenamefont {Lin}, \citenamefont {Guo}, \citenamefont
  {Zhang},\ and\ \citenamefont {He}}]{Liang2018_CNN}%
  \BibitemOpen
  \bibfield  {author} {\bibinfo {author} {\bibfnamefont {X.}~\bibnamefont
  {Liang}}, \bibinfo {author} {\bibfnamefont {W.-Y.}\ \bibnamefont {Liu}},
  \bibinfo {author} {\bibfnamefont {P.-Z.}\ \bibnamefont {Lin}}, \bibinfo
  {author} {\bibfnamefont {G.-C.}\ \bibnamefont {Guo}}, \bibinfo {author}
  {\bibfnamefont {Y.-S.}\ \bibnamefont {Zhang}},\ and\ \bibinfo {author}
  {\bibfnamefont {L.}~\bibnamefont {He}},\ }\bibfield  {title} {\bibinfo
  {title} {Solving frustrated quantum many-particle models with convolutional
  neural networks},\ }\href {https://doi.org/10.1103/PhysRevB.98.104426}
  {\bibfield  {journal} {\bibinfo  {journal} {Phys. Rev. B}\ }\textbf {\bibinfo
  {volume} {98}},\ \bibinfo {pages} {104426} (\bibinfo {year}
  {2018})}\BibitemShut {NoStop}%
\bibitem [{\citenamefont {Hibat-Allah}\ \emph {et~al.}(2020)\citenamefont
  {Hibat-Allah}, \citenamefont {Ganahl}, \citenamefont {Hayward}, \citenamefont
  {Melko},\ and\ \citenamefont {Carrasquilla}}]{HibatAllah2020_RNN_NQS}%
  \BibitemOpen
  \bibfield  {author} {\bibinfo {author} {\bibfnamefont {M.}~\bibnamefont
  {Hibat-Allah}}, \bibinfo {author} {\bibfnamefont {M.}~\bibnamefont {Ganahl}},
  \bibinfo {author} {\bibfnamefont {L.~E.}\ \bibnamefont {Hayward}}, \bibinfo
  {author} {\bibfnamefont {R.~G.}\ \bibnamefont {Melko}},\ and\ \bibinfo
  {author} {\bibfnamefont {J.}~\bibnamefont {Carrasquilla}},\ }\bibfield
  {title} {\bibinfo {title} {Recurrent neural network wave functions},\ }\href
  {https://doi.org/10.1103/PhysRevResearch.2.023358} {\bibfield  {journal}
  {\bibinfo  {journal} {Phys. Rev. Res.}\ }\textbf {\bibinfo {volume} {2}},\
  \bibinfo {pages} {023358} (\bibinfo {year} {2020})}\BibitemShut {NoStop}%
\bibitem [{\citenamefont {Szab\'o}\ and\ \citenamefont
  {Castelnovo}(2020)}]{Claudio2020_NQS_SignProblem}%
  \BibitemOpen
  \bibfield  {author} {\bibinfo {author} {\bibfnamefont {A.}~\bibnamefont
  {Szab\'o}}\ and\ \bibinfo {author} {\bibfnamefont {C.}~\bibnamefont
  {Castelnovo}},\ }\bibfield  {title} {\bibinfo {title} {Neural network wave
  functions and the sign problem},\ }\href
  {https://doi.org/10.1103/PhysRevResearch.2.033075} {\bibfield  {journal}
  {\bibinfo  {journal} {Phys. Rev. Res.}\ }\textbf {\bibinfo {volume} {2}},\
  \bibinfo {pages} {033075} (\bibinfo {year} {2020})}\BibitemShut {NoStop}%
\bibitem [{\citenamefont {Liang}\ \emph {et~al.}(2021)\citenamefont {Liang},
  \citenamefont {Dong},\ and\ \citenamefont {He}}]{Liang2021_CNN+PEPS}%
  \BibitemOpen
  \bibfield  {author} {\bibinfo {author} {\bibfnamefont {X.}~\bibnamefont
  {Liang}}, \bibinfo {author} {\bibfnamefont {S.-J.}\ \bibnamefont {Dong}},\
  and\ \bibinfo {author} {\bibfnamefont {L.}~\bibnamefont {He}},\ }\bibfield
  {title} {\bibinfo {title} {Hybrid convolutional neural network and projected
  entangled pair states wave functions for quantum many-particle states},\
  }\href {https://doi.org/10.1103/PhysRevB.103.035138} {\bibfield  {journal}
  {\bibinfo  {journal} {Phys. Rev. B}\ }\textbf {\bibinfo {volume} {103}},\
  \bibinfo {pages} {035138} (\bibinfo {year} {2021})}\BibitemShut {NoStop}%
\bibitem [{\citenamefont {Lin}\ and\ \citenamefont
  {Pollmann}(2022)}]{Lin2022_optimize_scale}%
  \BibitemOpen
  \bibfield  {author} {\bibinfo {author} {\bibfnamefont {S.-H.}\ \bibnamefont
  {Lin}}\ and\ \bibinfo {author} {\bibfnamefont {F.}~\bibnamefont {Pollmann}},\
  }\bibfield  {title} {\bibinfo {title} {Scaling of neural-network quantum
  states for time evolution},\ }\href
  {https://doi.org/https://doi.org/10.1002/pssb.202100172} {\bibfield
  {journal} {\bibinfo  {journal} {physica status solidi (b)}\ }\textbf
  {\bibinfo {volume} {259}},\ \bibinfo {pages} {2100172} (\bibinfo {year}
  {2022})}\BibitemShut {NoStop}%
\bibitem [{\citenamefont {Chen}\ and\ \citenamefont
  {Heyl}(2023)}]{chen2023_EfficientOptimization}%
  \BibitemOpen
  \bibfield  {author} {\bibinfo {author} {\bibfnamefont {A.}~\bibnamefont
  {Chen}}\ and\ \bibinfo {author} {\bibfnamefont {M.}~\bibnamefont {Heyl}},\
  }\bibfield  {title} {\bibinfo {title} {Efficient optimization of deep neural
  quantum states toward machine precision},\ }\href
  {https://doi.org/10.48550/arXiv.2302.01941} {\bibfield  {journal} {\bibinfo
  {journal} {arXiv:2302.01941}\ } (\bibinfo {year} {2023})}\BibitemShut
  {NoStop}%
\bibitem [{\citenamefont {Nomura}\ \emph {et~al.}(2017)\citenamefont {Nomura},
  \citenamefont {Darmawan}, \citenamefont {Yamaji},\ and\ \citenamefont
  {Imada}}]{Nomura2017_RBM+PP}%
  \BibitemOpen
  \bibfield  {author} {\bibinfo {author} {\bibfnamefont {Y.}~\bibnamefont
  {Nomura}}, \bibinfo {author} {\bibfnamefont {A.~S.}\ \bibnamefont
  {Darmawan}}, \bibinfo {author} {\bibfnamefont {Y.}~\bibnamefont {Yamaji}},\
  and\ \bibinfo {author} {\bibfnamefont {M.}~\bibnamefont {Imada}},\ }\bibfield
   {title} {\bibinfo {title} {Restricted boltzmann machine learning for solving
  strongly correlated quantum systems},\ }\href
  {https://doi.org/10.1103/PhysRevB.96.205152} {\bibfield  {journal} {\bibinfo
  {journal} {Phys. Rev. B}\ }\textbf {\bibinfo {volume} {96}},\ \bibinfo
  {pages} {205152} (\bibinfo {year} {2017})}\BibitemShut {NoStop}%
\bibitem [{\citenamefont {Nomura}\ and\ \citenamefont
  {Imada}(2021)}]{Nomura2021_RBM+PP}%
  \BibitemOpen
  \bibfield  {author} {\bibinfo {author} {\bibfnamefont {Y.}~\bibnamefont
  {Nomura}}\ and\ \bibinfo {author} {\bibfnamefont {M.}~\bibnamefont {Imada}},\
  }\bibfield  {title} {\bibinfo {title} {Dirac-type nodal spin liquid revealed
  by refined quantum many-body solver using neural-network wave function,
  correlation ratio, and level spectroscopy},\ }\href
  {https://doi.org/10.1103/PhysRevX.11.031034} {\bibfield  {journal} {\bibinfo
  {journal} {Phys. Rev. X}\ }\textbf {\bibinfo {volume} {11}},\ \bibinfo
  {pages} {031034} (\bibinfo {year} {2021})}\BibitemShut {NoStop}%
\bibitem [{\citenamefont {Pfau}\ \emph {et~al.}(2020)\citenamefont {Pfau},
  \citenamefont {Spencer}, \citenamefont {Matthews},\ and\ \citenamefont
  {Foulkes}}]{Pfau2020Ferminet}%
  \BibitemOpen
  \bibfield  {author} {\bibinfo {author} {\bibfnamefont {D.}~\bibnamefont
  {Pfau}}, \bibinfo {author} {\bibfnamefont {J.~S.}\ \bibnamefont {Spencer}},
  \bibinfo {author} {\bibfnamefont {A.~G. D.~G.}\ \bibnamefont {Matthews}},\
  and\ \bibinfo {author} {\bibfnamefont {W.~M.~C.}\ \bibnamefont {Foulkes}},\
  }\bibfield  {title} {\bibinfo {title} {Ab initio solution of the
  many-electron schr\"odinger equation with deep neural networks},\ }\href
  {https://doi.org/10.1103/PhysRevResearch.2.033429} {\bibfield  {journal}
  {\bibinfo  {journal} {Phys. Rev. Res.}\ }\textbf {\bibinfo {volume} {2}},\
  \bibinfo {pages} {033429} (\bibinfo {year} {2020})}\BibitemShut {NoStop}%
\bibitem [{\citenamefont {Lou}\ and\ \citenamefont
  {Sandvik}(2007)}]{Sandvik2007VBBasis}%
  \BibitemOpen
  \bibfield  {author} {\bibinfo {author} {\bibfnamefont {J.}~\bibnamefont
  {Lou}}\ and\ \bibinfo {author} {\bibfnamefont {A.~W.}\ \bibnamefont
  {Sandvik}},\ }\bibfield  {title} {\bibinfo {title} {Variational ground states
  of two-dimensional antiferromagnets in the valence bond basis},\ }\href
  {https://doi.org/10.1103/PhysRevB.76.104432} {\bibfield  {journal} {\bibinfo
  {journal} {Phys. Rev. B}\ }\textbf {\bibinfo {volume} {76}},\ \bibinfo
  {pages} {104432} (\bibinfo {year} {2007})}\BibitemShut {NoStop}%
\bibitem [{\citenamefont {Sandvik}\ and\ \citenamefont
  {Evertz}(2010)}]{Sandvik2010_VBBasis+LoopUpdate}%
  \BibitemOpen
  \bibfield  {author} {\bibinfo {author} {\bibfnamefont {A.~W.}\ \bibnamefont
  {Sandvik}}\ and\ \bibinfo {author} {\bibfnamefont {H.~G.}\ \bibnamefont
  {Evertz}},\ }\bibfield  {title} {\bibinfo {title} {Loop updates for
  variational and projector quantum monte carlo simulations in the valence-bond
  basis},\ }\href {https://doi.org/10.1103/PhysRevB.82.024407} {\bibfield
  {journal} {\bibinfo  {journal} {Phys. Rev. B}\ }\textbf {\bibinfo {volume}
  {82}},\ \bibinfo {pages} {024407} (\bibinfo {year} {2010})}\BibitemShut
  {NoStop}%
\bibitem [{\citenamefont {Hu}\ \emph {et~al.}(2013)\citenamefont {Hu},
  \citenamefont {Becca}, \citenamefont {Parola},\ and\ \citenamefont
  {Sorella}}]{Sorella2013_VMC}%
  \BibitemOpen
  \bibfield  {author} {\bibinfo {author} {\bibfnamefont {W.-J.}\ \bibnamefont
  {Hu}}, \bibinfo {author} {\bibfnamefont {F.}~\bibnamefont {Becca}}, \bibinfo
  {author} {\bibfnamefont {A.}~\bibnamefont {Parola}},\ and\ \bibinfo {author}
  {\bibfnamefont {S.}~\bibnamefont {Sorella}},\ }\bibfield  {title} {\bibinfo
  {title} {Direct evidence for a gapless ${Z}_{2}$ spin liquid by frustrating
  n\'eel antiferromagnetism},\ }\href
  {https://doi.org/10.1103/PhysRevB.88.060402} {\bibfield  {journal} {\bibinfo
  {journal} {Phys. Rev. B}\ }\textbf {\bibinfo {volume} {88}},\ \bibinfo
  {pages} {060402(R)} (\bibinfo {year} {2013})}\BibitemShut {NoStop}%
\bibitem [{\citenamefont {Sandvik}\ and\ \citenamefont
  {Vidal}(2007)}]{Sandvik2007_VMC+TNS}%
  \BibitemOpen
  \bibfield  {author} {\bibinfo {author} {\bibfnamefont {A.~W.}\ \bibnamefont
  {Sandvik}}\ and\ \bibinfo {author} {\bibfnamefont {G.}~\bibnamefont
  {Vidal}},\ }\bibfield  {title} {\bibinfo {title} {Variational quantum monte
  carlo simulations with tensor-network states},\ }\href
  {https://doi.org/10.1103/PhysRevLett.99.220602} {\bibfield  {journal}
  {\bibinfo  {journal} {Phys. Rev. Lett.}\ }\textbf {\bibinfo {volume} {99}},\
  \bibinfo {pages} {220602} (\bibinfo {year} {2007})}\BibitemShut {NoStop}%
\bibitem [{\citenamefont {Tagliacozzo}\ \emph {et~al.}(2009)\citenamefont
  {Tagliacozzo}, \citenamefont {Evenbly},\ and\ \citenamefont
  {Vidal}}]{Tagliacozzo2009_TreeTensorNetwork}%
  \BibitemOpen
  \bibfield  {author} {\bibinfo {author} {\bibfnamefont {L.}~\bibnamefont
  {Tagliacozzo}}, \bibinfo {author} {\bibfnamefont {G.}~\bibnamefont
  {Evenbly}},\ and\ \bibinfo {author} {\bibfnamefont {G.}~\bibnamefont
  {Vidal}},\ }\bibfield  {title} {\bibinfo {title} {Simulation of
  two-dimensional quantum systems using a tree tensor network that exploits the
  entropic area law},\ }\href {https://doi.org/10.1103/PhysRevB.80.235127}
  {\bibfield  {journal} {\bibinfo  {journal} {Phys. Rev. B}\ }\textbf {\bibinfo
  {volume} {80}},\ \bibinfo {pages} {235127} (\bibinfo {year}
  {2009})}\BibitemShut {NoStop}%
\bibitem [{\citenamefont {Gong}\ \emph {et~al.}(2014)\citenamefont {Gong},
  \citenamefont {Zhu}, \citenamefont {Sheng}, \citenamefont {Motrunich},\ and\
  \citenamefont {Fisher}}]{GongShouShu2014_DMRG}%
  \BibitemOpen
  \bibfield  {author} {\bibinfo {author} {\bibfnamefont {S.-S.}\ \bibnamefont
  {Gong}}, \bibinfo {author} {\bibfnamefont {W.}~\bibnamefont {Zhu}}, \bibinfo
  {author} {\bibfnamefont {D.~N.}\ \bibnamefont {Sheng}}, \bibinfo {author}
  {\bibfnamefont {O.~I.}\ \bibnamefont {Motrunich}},\ and\ \bibinfo {author}
  {\bibfnamefont {M.~P.~A.}\ \bibnamefont {Fisher}},\ }\bibfield  {title}
  {\bibinfo {title} {Plaquette ordered phase and quantum phase diagram in the
  spin-$\frac{1}{2}$ ${J}_{1}\text{\ensuremath{-}}{J}_{2}$ square heisenberg
  model},\ }\href {https://doi.org/10.1103/PhysRevLett.113.027201} {\bibfield
  {journal} {\bibinfo  {journal} {Phys. Rev. Lett.}\ }\textbf {\bibinfo
  {volume} {113}},\ \bibinfo {pages} {027201} (\bibinfo {year}
  {2014})}\BibitemShut {NoStop}%
\bibitem [{\citenamefont {Stoudenmire}\ and\ \citenamefont
  {White}(2012)}]{Stoudenmire2012_2DsystemDMRG}%
  \BibitemOpen
  \bibfield  {author} {\bibinfo {author} {\bibfnamefont {E.}~\bibnamefont
  {Stoudenmire}}\ and\ \bibinfo {author} {\bibfnamefont {S.~R.}\ \bibnamefont
  {White}},\ }\bibfield  {title} {\bibinfo {title} {Studying two-dimensional
  systems with the density matrix renormalization group},\ }\href
  {https://doi.org/10.1146/annurev-conmatphys-020911-125018} {\bibfield
  {journal} {\bibinfo  {journal} {Annual Review of Condensed Matter Physics}\
  }\textbf {\bibinfo {volume} {3}},\ \bibinfo {pages} {111} (\bibinfo {year}
  {2012})}\BibitemShut {NoStop}%
\bibitem [{\citenamefont {Wang}\ and\ \citenamefont
  {Sandvik}(2018)}]{Sandvik2018_DMRG}%
  \BibitemOpen
  \bibfield  {author} {\bibinfo {author} {\bibfnamefont {L.}~\bibnamefont
  {Wang}}\ and\ \bibinfo {author} {\bibfnamefont {A.~W.}\ \bibnamefont
  {Sandvik}},\ }\bibfield  {title} {\bibinfo {title} {Critical level crossings
  and gapless spin liquid in the square-lattice spin-$1/2$
  ${J}_{1}\ensuremath{-}{J}_{2}$ heisenberg antiferromagnet},\ }\href
  {https://doi.org/10.1103/PhysRevLett.121.107202} {\bibfield  {journal}
  {\bibinfo  {journal} {Phys. Rev. Lett.}\ }\textbf {\bibinfo {volume} {121}},\
  \bibinfo {pages} {107202} (\bibinfo {year} {2018})}\BibitemShut {NoStop}%
\bibitem [{\citenamefont {Sandvik}(1997)}]{Sandvik1997SSEQMC}%
  \BibitemOpen
  \bibfield  {author} {\bibinfo {author} {\bibfnamefont {A.~W.}\ \bibnamefont
  {Sandvik}},\ }\bibfield  {title} {\bibinfo {title} {Finite-size scaling of
  the ground-state parameters of the two-dimensional heisenberg model},\ }\href
  {https://doi.org/10.1103/PhysRevB.56.11678} {\bibfield  {journal} {\bibinfo
  {journal} {Phys. Rev. B}\ }\textbf {\bibinfo {volume} {56}},\ \bibinfo
  {pages} {11678} (\bibinfo {year} {1997})}\BibitemShut {NoStop}%
\bibitem [{\citenamefont {Zhao}\ \emph {et~al.}(2023)\citenamefont {Zhao},
  \citenamefont {Zhou}, \citenamefont {Li}, \citenamefont {Tang}, \citenamefont
  {Wang}, \citenamefont {Hou}, \citenamefont {Min}, \citenamefont {Zhang},
  \citenamefont {Zhang}, \citenamefont {Dong}, \citenamefont {Du},
  \citenamefont {Yang}, \citenamefont {Chen}, \citenamefont {Chen},
  \citenamefont {Jiang}, \citenamefont {Ren}, \citenamefont {Li}, \citenamefont
  {Tang}, \citenamefont {Liu}, \citenamefont {Liu}, \citenamefont {Nie},\ and\
  \citenamefont {Wen}}]{zhao2023LLMSurvey}%
  \BibitemOpen
  \bibfield  {author} {\bibinfo {author} {\bibfnamefont {W.~X.}\ \bibnamefont
  {Zhao}}, \bibinfo {author} {\bibfnamefont {K.}~\bibnamefont {Zhou}}, \bibinfo
  {author} {\bibfnamefont {J.}~\bibnamefont {Li}}, \bibinfo {author}
  {\bibfnamefont {T.}~\bibnamefont {Tang}}, \bibinfo {author} {\bibfnamefont
  {X.}~\bibnamefont {Wang}}, \bibinfo {author} {\bibfnamefont {Y.}~\bibnamefont
  {Hou}}, \bibinfo {author} {\bibfnamefont {Y.}~\bibnamefont {Min}}, \bibinfo
  {author} {\bibfnamefont {B.}~\bibnamefont {Zhang}}, \bibinfo {author}
  {\bibfnamefont {J.}~\bibnamefont {Zhang}}, \bibinfo {author} {\bibfnamefont
  {Z.}~\bibnamefont {Dong}}, \bibinfo {author} {\bibfnamefont {Y.}~\bibnamefont
  {Du}}, \bibinfo {author} {\bibfnamefont {C.}~\bibnamefont {Yang}}, \bibinfo
  {author} {\bibfnamefont {Y.}~\bibnamefont {Chen}}, \bibinfo {author}
  {\bibfnamefont {Z.}~\bibnamefont {Chen}}, \bibinfo {author} {\bibfnamefont
  {J.}~\bibnamefont {Jiang}}, \bibinfo {author} {\bibfnamefont
  {R.}~\bibnamefont {Ren}}, \bibinfo {author} {\bibfnamefont {Y.}~\bibnamefont
  {Li}}, \bibinfo {author} {\bibfnamefont {X.}~\bibnamefont {Tang}}, \bibinfo
  {author} {\bibfnamefont {Z.}~\bibnamefont {Liu}}, \bibinfo {author}
  {\bibfnamefont {P.}~\bibnamefont {Liu}}, \bibinfo {author} {\bibfnamefont
  {J.-Y.}\ \bibnamefont {Nie}},\ and\ \bibinfo {author} {\bibfnamefont {J.-R.}\
  \bibnamefont {Wen}},\ }\bibfield  {title} {\bibinfo {title} {A survey of
  large language models},\ }\href {https://doi.org/10.48550/arXiv.2303.18223}
  {\bibfield  {journal} {\bibinfo  {journal} {arXiv:2303.18223}\ } (\bibinfo
  {year} {2023})}\BibitemShut {NoStop}%
\bibitem [{\citenamefont {Wei}\ \emph {et~al.}(2022)\citenamefont {Wei},
  \citenamefont {Tay}, \citenamefont {Bommasani}, \citenamefont {Raffel},
  \citenamefont {Zoph}, \citenamefont {Borgeaud}, \citenamefont {Yogatama},
  \citenamefont {Bosma}, \citenamefont {Zhou}, \citenamefont {Metzler},
  \citenamefont {Chi}, \citenamefont {Hashimoto}, \citenamefont {Vinyals},
  \citenamefont {Liang}, \citenamefont {Dean},\ and\ \citenamefont
  {Fedus}}]{Wei2022EmergentAbilities}%
  \BibitemOpen
  \bibfield  {author} {\bibinfo {author} {\bibfnamefont {J.}~\bibnamefont
  {Wei}}, \bibinfo {author} {\bibfnamefont {Y.}~\bibnamefont {Tay}}, \bibinfo
  {author} {\bibfnamefont {R.}~\bibnamefont {Bommasani}}, \bibinfo {author}
  {\bibfnamefont {C.}~\bibnamefont {Raffel}}, \bibinfo {author} {\bibfnamefont
  {B.}~\bibnamefont {Zoph}}, \bibinfo {author} {\bibfnamefont {S.}~\bibnamefont
  {Borgeaud}}, \bibinfo {author} {\bibfnamefont {D.}~\bibnamefont {Yogatama}},
  \bibinfo {author} {\bibfnamefont {M.}~\bibnamefont {Bosma}}, \bibinfo
  {author} {\bibfnamefont {D.}~\bibnamefont {Zhou}}, \bibinfo {author}
  {\bibfnamefont {D.}~\bibnamefont {Metzler}}, \bibinfo {author} {\bibfnamefont
  {E.~H.}\ \bibnamefont {Chi}}, \bibinfo {author} {\bibfnamefont
  {T.}~\bibnamefont {Hashimoto}}, \bibinfo {author} {\bibfnamefont
  {O.}~\bibnamefont {Vinyals}}, \bibinfo {author} {\bibfnamefont
  {P.}~\bibnamefont {Liang}}, \bibinfo {author} {\bibfnamefont
  {J.}~\bibnamefont {Dean}},\ and\ \bibinfo {author} {\bibfnamefont
  {W.}~\bibnamefont {Fedus}},\ }\bibfield  {title} {\bibinfo {title} {Emergent
  abilities of large language models},\ }\href
  {https://doi.org/10.48550/arXiv.2206.07682} {\bibfield  {journal} {\bibinfo
  {journal} {arXiv:2206.07682}\ } (\bibinfo {year} {2022})}\BibitemShut
  {NoStop}%
\bibitem [{\citenamefont {Ruder}(2017)}]{ruder2017overview_GradientDescent}%
  \BibitemOpen
  \bibfield  {author} {\bibinfo {author} {\bibfnamefont {S.}~\bibnamefont
  {Ruder}},\ }\bibfield  {title} {\bibinfo {title} {An overview of gradient
  descent optimization algorithms},\ }\href
  {https://doi.org/10.48550/arXiv.1609.04747} {\bibfield  {journal} {\bibinfo
  {journal} {arXiv:1609.04747}\ } (\bibinfo {year} {2017})}\BibitemShut
  {NoStop}%
\bibitem [{\citenamefont {Vaswani}\ \emph {et~al.}(2017)\citenamefont
  {Vaswani}, \citenamefont {Shazeer}, \citenamefont {Parmar}, \citenamefont
  {Uszkoreit}, \citenamefont {Jones}, \citenamefont {Gomez}, \citenamefont
  {Kaiser},\ and\ \citenamefont {Polosukhin}}]{vaswani2023attention}%
  \BibitemOpen
  \bibfield  {author} {\bibinfo {author} {\bibfnamefont {A.}~\bibnamefont
  {Vaswani}}, \bibinfo {author} {\bibfnamefont {N.}~\bibnamefont {Shazeer}},
  \bibinfo {author} {\bibfnamefont {N.}~\bibnamefont {Parmar}}, \bibinfo
  {author} {\bibfnamefont {J.}~\bibnamefont {Uszkoreit}}, \bibinfo {author}
  {\bibfnamefont {L.}~\bibnamefont {Jones}}, \bibinfo {author} {\bibfnamefont
  {A.~N.}\ \bibnamefont {Gomez}}, \bibinfo {author} {\bibfnamefont
  {L.}~\bibnamefont {Kaiser}},\ and\ \bibinfo {author} {\bibfnamefont
  {I.}~\bibnamefont {Polosukhin}},\ }\bibfield  {title} {\bibinfo {title}
  {Attention is all you need},\ }\href
  {https://proceedings.neurips.cc/paper_files/paper/2017/file/3f5ee243547dee91fbd053c1c4a845aa-Paper.pdf}
  {\bibfield  {journal} {\bibinfo  {journal} {Adv. Neural Inf. Process. Syst.}\
  }\textbf {\bibinfo {volume} {30}},\ \bibinfo {pages} {5889} (\bibinfo {year}
  {2017})}\BibitemShut {NoStop}%
\bibitem [{\citenamefont {He}\ \emph {et~al.}(2016)\citenamefont {He},
  \citenamefont {Zhang}, \citenamefont {Ren},\ and\ \citenamefont
  {Sun}}]{he2015Resnet}%
  \BibitemOpen
  \bibfield  {author} {\bibinfo {author} {\bibfnamefont {K.}~\bibnamefont
  {He}}, \bibinfo {author} {\bibfnamefont {X.}~\bibnamefont {Zhang}}, \bibinfo
  {author} {\bibfnamefont {S.}~\bibnamefont {Ren}},\ and\ \bibinfo {author}
  {\bibfnamefont {J.}~\bibnamefont {Sun}},\ }\bibfield  {title} {\bibinfo
  {title} {Deep residual learning for image recognition},\ }in\ \href
  {https://doi.org/10.1109/CVPR.2016.90} {\emph {\bibinfo {booktitle}
  {Proceedings of the IEEE Conference on Computer Vision and Pattern
  Recognition (CVPR)}}}\ (\bibinfo  {publisher} {IEEE, New York},\ \bibinfo
  {year} {2016})\ pp.\ \bibinfo {pages} {770--778}\BibitemShut {NoStop}%
\bibitem [{\citenamefont {Park}\ and\ \citenamefont
  {Kastoryano}(2022)}]{Park2022_Frustrated_RBM}%
  \BibitemOpen
  \bibfield  {author} {\bibinfo {author} {\bibfnamefont {C.-Y.}\ \bibnamefont
  {Park}}\ and\ \bibinfo {author} {\bibfnamefont {M.~J.}\ \bibnamefont
  {Kastoryano}},\ }\bibfield  {title} {\bibinfo {title} {Expressive power of
  complex-valued restricted boltzmann machines for solving nonstoquastic
  hamiltonians},\ }\href {https://doi.org/10.1103/PhysRevB.106.134437}
  {\bibfield  {journal} {\bibinfo  {journal} {Phys. Rev. B}\ }\textbf {\bibinfo
  {volume} {106}},\ \bibinfo {pages} {134437} (\bibinfo {year}
  {2022})}\BibitemShut {NoStop}%
\bibitem [{\citenamefont {Chen}\ \emph {et~al.}(2022)\citenamefont {Chen},
  \citenamefont {Choo}, \citenamefont {Astrakhantsev},\ and\ \citenamefont
  {Neupert}}]{Chen2022_SignStructure-NeuralNetwork}%
  \BibitemOpen
  \bibfield  {author} {\bibinfo {author} {\bibfnamefont {A.}~\bibnamefont
  {Chen}}, \bibinfo {author} {\bibfnamefont {K.}~\bibnamefont {Choo}}, \bibinfo
  {author} {\bibfnamefont {N.}~\bibnamefont {Astrakhantsev}},\ and\ \bibinfo
  {author} {\bibfnamefont {T.}~\bibnamefont {Neupert}},\ }\bibfield  {title}
  {\bibinfo {title} {Neural network evolution strategy for solving quantum sign
  structures},\ }\href {https://doi.org/10.1103/PhysRevResearch.4.L022026}
  {\bibfield  {journal} {\bibinfo  {journal} {Phys. Rev. Res.}\ }\textbf
  {\bibinfo {volume} {4}},\ \bibinfo {pages} {L022026} (\bibinfo {year}
  {2022})}\BibitemShut {NoStop}%
\bibitem [{\citenamefont {Westerhout}\ \emph {et~al.}(2023)\citenamefont
  {Westerhout}, \citenamefont {Katsnelson},\ and\ \citenamefont
  {Bagrov}}]{westerhout2023SignStructure}%
  \BibitemOpen
  \bibfield  {author} {\bibinfo {author} {\bibfnamefont {T.}~\bibnamefont
  {Westerhout}}, \bibinfo {author} {\bibfnamefont {M.~I.}\ \bibnamefont
  {Katsnelson}},\ and\ \bibinfo {author} {\bibfnamefont {A.~A.}\ \bibnamefont
  {Bagrov}},\ }\bibfield  {title} {\bibinfo {title} {Many-body quantum sign
  structures as non-glassy ising models},\ }\href
  {https://doi.org/10.1038/s42005-023-01388-6} {\bibfield  {journal} {\bibinfo
  {journal} {Commun. Phys.}\ }\textbf {\bibinfo {volume} {6}},\ \bibinfo
  {pages} {275} (\bibinfo {year} {2023})}\BibitemShut {NoStop}%
\bibitem [{\citenamefont {Westerhout}\ \emph {et~al.}(2020)\citenamefont
  {Westerhout}, \citenamefont {Astrakhantsev}, \citenamefont {Tikhonov},
  \citenamefont {Katsnelson},\ and\ \citenamefont
  {Bagrov}}]{Westerhout2020_difference_Sign_amp}%
  \BibitemOpen
  \bibfield  {author} {\bibinfo {author} {\bibfnamefont {T.}~\bibnamefont
  {Westerhout}}, \bibinfo {author} {\bibfnamefont {N.}~\bibnamefont
  {Astrakhantsev}}, \bibinfo {author} {\bibfnamefont {K.~S.}\ \bibnamefont
  {Tikhonov}}, \bibinfo {author} {\bibfnamefont {M.~I.}\ \bibnamefont
  {Katsnelson}},\ and\ \bibinfo {author} {\bibfnamefont {A.~A.}\ \bibnamefont
  {Bagrov}},\ }\bibfield  {title} {\bibinfo {title} {Generalization properties
  of neural network approximations to frustrated magnet ground states},\ }\href
  {https://doi.org/10.1038/s41467-020-15402-w} {\bibfield  {journal} {\bibinfo
  {journal} {Nat. Commun.}\ }\textbf {\bibinfo {volume} {11}},\ \bibinfo
  {pages} {1593} (\bibinfo {year} {2020})}\BibitemShut {NoStop}%
\bibitem [{\citenamefont {Hornik}\ \emph {et~al.}(1989)\citenamefont {Hornik},
  \citenamefont {Stinchcombe},\ and\ \citenamefont {White}}]{Horink1989UAT}%
  \BibitemOpen
  \bibfield  {author} {\bibinfo {author} {\bibfnamefont {K.}~\bibnamefont
  {Hornik}}, \bibinfo {author} {\bibfnamefont {M.}~\bibnamefont
  {Stinchcombe}},\ and\ \bibinfo {author} {\bibfnamefont {H.}~\bibnamefont
  {White}},\ }\bibfield  {title} {\bibinfo {title} {Multilayer feedforward
  networks are universal approximators},\ }\href
  {https://doi.org/https://doi.org/10.1016/0893-6080(89)90020-8} {\bibfield
  {journal} {\bibinfo  {journal} {Neural Networks}\ }\textbf {\bibinfo {volume}
  {2}},\ \bibinfo {pages} {359} (\bibinfo {year} {1989})}\BibitemShut {NoStop}%
\bibitem [{\citenamefont {Cybenko}(1989)}]{Cybenko1989ApproximationBS}%
  \BibitemOpen
  \bibfield  {author} {\bibinfo {author} {\bibfnamefont {G.~V.}\ \bibnamefont
  {Cybenko}},\ }\bibfield  {title} {\bibinfo {title} {Approximation by
  superpositions of a sigmoidal function},\ }\href
  {https://api.semanticscholar.org/CorpusID:3958369} {\bibfield  {journal}
  {\bibinfo  {journal} {Mathematics of Control, Signals and Systems}\ }\textbf
  {\bibinfo {volume} {2}},\ \bibinfo {pages} {303} (\bibinfo {year}
  {1989})}\BibitemShut {NoStop}%
\bibitem [{\citenamefont {Zhang}\ \emph
  {et~al.}(2023{\natexlab{b}})\citenamefont {Zhang}, \citenamefont {Wan},\ and\
  \citenamefont {Yao}}]{zhang2023ADVMC}%
  \BibitemOpen
  \bibfield  {author} {\bibinfo {author} {\bibfnamefont {S.-X.}\ \bibnamefont
  {Zhang}}, \bibinfo {author} {\bibfnamefont {Z.-Q.}\ \bibnamefont {Wan}},\
  and\ \bibinfo {author} {\bibfnamefont {H.}~\bibnamefont {Yao}},\ }\bibfield
  {title} {\bibinfo {title} {Automatic differentiable monte carlo: Theory and
  application},\ }\href {https://doi.org/10.1103/PhysRevResearch.5.033041}
  {\bibfield  {journal} {\bibinfo  {journal} {Phys. Rev. Res.}\ }\textbf
  {\bibinfo {volume} {5}},\ \bibinfo {pages} {033041} (\bibinfo {year}
  {2023}{\natexlab{b}})}\BibitemShut {NoStop}%
\bibitem [{\citenamefont {Sorella}\ \emph {et~al.}(2007)\citenamefont
  {Sorella}, \citenamefont {Casula},\ and\ \citenamefont
  {Rocca}}]{Sorella2007SR}%
  \BibitemOpen
  \bibfield  {author} {\bibinfo {author} {\bibfnamefont {S.}~\bibnamefont
  {Sorella}}, \bibinfo {author} {\bibfnamefont {M.}~\bibnamefont {Casula}},\
  and\ \bibinfo {author} {\bibfnamefont {D.}~\bibnamefont {Rocca}},\ }\bibfield
   {title} {\bibinfo {title} {Weak binding between two aromatic rings: Feeling
  the van der waals attraction by quantum monte carlo methods},\ }\bibfield
  {journal} {\bibinfo  {journal} {JOURNAL OF CHEMICAL PHYSICS}\ }\textbf
  {\bibinfo {volume} {127}},\ \href {https://doi.org/10.1063/1.2746035}
  {10.1063/1.2746035} (\bibinfo {year} {2007})\BibitemShut {NoStop}%
\bibitem [{\citenamefont {Kingma}\ and\ \citenamefont
  {Ba}(2017)}]{Kingma2015Adam}%
  \BibitemOpen
  \bibfield  {author} {\bibinfo {author} {\bibfnamefont {D.~P.}\ \bibnamefont
  {Kingma}}\ and\ \bibinfo {author} {\bibfnamefont {J.}~\bibnamefont {Ba}},\
  }\bibfield  {title} {\bibinfo {title} {Adam: A method for stochastic
  optimization},\ }\href {https://doi.org/10.48550/arXiv.1412.6980} {\bibfield
  {journal} {\bibinfo  {journal} {arXiv:1412.6980}\ } (\bibinfo {year}
  {2017})}\BibitemShut {NoStop}%
\bibitem [{\citenamefont {He}\ \emph {et~al.}(2015)\citenamefont {He},
  \citenamefont {Zhang}, \citenamefont {Ren},\ and\ \citenamefont
  {Sun}}]{he2015KaimingInitilization}%
  \BibitemOpen
  \bibfield  {author} {\bibinfo {author} {\bibfnamefont {K.}~\bibnamefont
  {He}}, \bibinfo {author} {\bibfnamefont {X.}~\bibnamefont {Zhang}}, \bibinfo
  {author} {\bibfnamefont {S.}~\bibnamefont {Ren}},\ and\ \bibinfo {author}
  {\bibfnamefont {J.}~\bibnamefont {Sun}},\ }\bibfield  {title} {\bibinfo
  {title} {Delving deep into rectifiers: Surpassing human-level performance on
  imagenet classification},\ }in\ \href@noop {} {\emph {\bibinfo {booktitle}
  {Proceedings of the IEEE International Conference on Computer Vision}}}\
  (\bibinfo  {publisher} {IEEE, Piscataway},\ \bibinfo {year} {2015})\ pp.\
  \bibinfo {pages} {1026--1034}\BibitemShut {NoStop}%
\bibitem [{\citenamefont {Marshall}(1955)}]{1955MarshallSign}%
  \BibitemOpen
  \bibfield  {author} {\bibinfo {author} {\bibfnamefont {W.}~\bibnamefont
  {Marshall}},\ }\bibfield  {title} {\bibinfo {title} {Antiferromagnetism},\
  }\href@noop {} {\bibfield  {journal} {\bibinfo  {journal} {Proc. R. Soc.
  Lond. A}\ }\textbf {\bibinfo {volume} {232}},\ \bibinfo {pages} {48}
  (\bibinfo {year} {1955})}\BibitemShut {NoStop}%
\bibitem [{\citenamefont {Bl\"ote}\ and\ \citenamefont
  {Deng}(2002)}]{DengYoujin2002PRETFIM}%
  \BibitemOpen
  \bibfield  {author} {\bibinfo {author} {\bibfnamefont {H.~W.~J.}\
  \bibnamefont {Bl\"ote}}\ and\ \bibinfo {author} {\bibfnamefont
  {Y.}~\bibnamefont {Deng}},\ }\bibfield  {title} {\bibinfo {title} {Cluster
  monte carlo simulation of the transverse ising model},\ }\href
  {https://doi.org/10.1103/PhysRevE.66.066110} {\bibfield  {journal} {\bibinfo
  {journal} {Phys. Rev. E}\ }\textbf {\bibinfo {volume} {66}},\ \bibinfo
  {pages} {066110} (\bibinfo {year} {2002})}\BibitemShut {NoStop}%
\bibitem [{\citenamefont {Qian}\ and\ \citenamefont
  {Qin}(2022)}]{QinMingpu2022_TreeTensorNetwork-MERA}%
  \BibitemOpen
  \bibfield  {author} {\bibinfo {author} {\bibfnamefont {X.}~\bibnamefont
  {Qian}}\ and\ \bibinfo {author} {\bibfnamefont {M.}~\bibnamefont {Qin}},\
  }\bibfield  {title} {\bibinfo {title} {From tree tensor network to multiscale
  entanglement renormalization ansatz},\ }\href
  {https://doi.org/10.1103/PhysRevB.105.205102} {\bibfield  {journal} {\bibinfo
   {journal} {Phys. Rev. B}\ }\textbf {\bibinfo {volume} {105}},\ \bibinfo
  {pages} {205102} (\bibinfo {year} {2022})}\BibitemShut {NoStop}%
\bibitem [{\citenamefont {Qian}\ and\ \citenamefont
  {Qin}(2023)}]{QinMingpu2023_DMRG-FAMPS}%
  \BibitemOpen
  \bibfield  {author} {\bibinfo {author} {\bibfnamefont {X.}~\bibnamefont
  {Qian}}\ and\ \bibinfo {author} {\bibfnamefont {M.}~\bibnamefont {Qin}},\
  }\bibfield  {title} {\bibinfo {title} {Augmenting density matrix
  renormalization group with disentanglers},\ }\href
  {https://doi.org/10.1088/0256-307x/40/5/057102} {\bibfield  {journal}
  {\bibinfo  {journal} {Chinese Physics Letters}\ }\textbf {\bibinfo {volume}
  {40}},\ \bibinfo {pages} {057102} (\bibinfo {year} {2023})}\BibitemShut
  {NoStop}%
\bibitem [{\citenamefont {Choo}\ \emph {et~al.}(2019)\citenamefont {Choo},
  \citenamefont {Neupert},\ and\ \citenamefont
  {Carleo}}]{Choo2019_CNN_ComplexValued}%
  \BibitemOpen
  \bibfield  {author} {\bibinfo {author} {\bibfnamefont {K.}~\bibnamefont
  {Choo}}, \bibinfo {author} {\bibfnamefont {T.}~\bibnamefont {Neupert}},\ and\
  \bibinfo {author} {\bibfnamefont {G.}~\bibnamefont {Carleo}},\ }\bibfield
  {title} {\bibinfo {title} {Two-dimensional frustrated
  ${J}_{1}\text{\ensuremath{-}}{J}_{2}$ model studied with neural network
  quantum states},\ }\href {https://doi.org/10.1103/PhysRevB.100.125124}
  {\bibfield  {journal} {\bibinfo  {journal} {Phys. Rev. B}\ }\textbf {\bibinfo
  {volume} {100}},\ \bibinfo {pages} {125124} (\bibinfo {year}
  {2019})}\BibitemShut {NoStop}%
\bibitem [{\citenamefont {Rende}\ \emph {et~al.}(2023)\citenamefont {Rende},
  \citenamefont {Viteritti}, \citenamefont {Bardone}, \citenamefont {Becca},\
  and\ \citenamefont {Goldt}}]{rende2023DeepViT}%
  \BibitemOpen
  \bibfield  {author} {\bibinfo {author} {\bibfnamefont {R.}~\bibnamefont
  {Rende}}, \bibinfo {author} {\bibfnamefont {L.~L.}\ \bibnamefont
  {Viteritti}}, \bibinfo {author} {\bibfnamefont {L.}~\bibnamefont {Bardone}},
  \bibinfo {author} {\bibfnamefont {F.}~\bibnamefont {Becca}},\ and\ \bibinfo
  {author} {\bibfnamefont {S.}~\bibnamefont {Goldt}},\ }\bibfield  {title}
  {\bibinfo {title} {A simple linear algebra identity to optimize large-scale
  neural network quantum states},\ }\href
  {https://doi.org/10.48550/arXiv.2310.05715} {\bibfield  {journal} {\bibinfo
  {journal} {arXiv:2310.05715}\ } (\bibinfo {year} {2023})}\BibitemShut
  {NoStop}%
\bibitem [{\citenamefont {Voigt}\ \emph {et~al.}(1997)\citenamefont {Voigt},
  \citenamefont {Richter},\ and\ \citenamefont
  {Ivanov}}]{Ivanov1997_MSR_exact_in_J1J2}%
  \BibitemOpen
  \bibfield  {author} {\bibinfo {author} {\bibfnamefont {A.}~\bibnamefont
  {Voigt}}, \bibinfo {author} {\bibfnamefont {J.}~\bibnamefont {Richter}},\
  and\ \bibinfo {author} {\bibfnamefont {N.~B.}\ \bibnamefont {Ivanov}},\
  }\bibfield  {title} {\bibinfo {title} {Marshall-peierls sign rule for excited
  states of the frustrated j1-j2 heisenberg antiferromagnet},\ }\href
  {https://doi.org/https://doi.org/10.1016/S0378-4371(97)00330-0} {\bibfield
  {journal} {\bibinfo  {journal} {Physica A: Statistical Mechanics and its
  Applications}\ }\textbf {\bibinfo {volume} {245}},\ \bibinfo {pages} {269}
  (\bibinfo {year} {1997})}\BibitemShut {NoStop}%
\bibitem [{\citenamefont {Fu}\ \emph {et~al.}(2024)\citenamefont {Fu},
  \citenamefont {Zhang}, \citenamefont {Zhang}, \citenamefont {Ling},
  \citenamefont {Xu},\ and\ \citenamefont {Ji}}]{fu2022latticeCNN}%
  \BibitemOpen
  \bibfield  {author} {\bibinfo {author} {\bibfnamefont {C.}~\bibnamefont
  {Fu}}, \bibinfo {author} {\bibfnamefont {X.}~\bibnamefont {Zhang}}, \bibinfo
  {author} {\bibfnamefont {H.}~\bibnamefont {Zhang}}, \bibinfo {author}
  {\bibfnamefont {H.}~\bibnamefont {Ling}}, \bibinfo {author} {\bibfnamefont
  {S.}~\bibnamefont {Xu}},\ and\ \bibinfo {author} {\bibfnamefont
  {S.}~\bibnamefont {Ji}},\ }\bibfield  {title} {\bibinfo {title} {Lattice
  convolutional networks for learning ground states of quantum many-body
  systems},\ }in\ \href {https://doi.org/10.1137/1.9781611978032.57} {\emph
  {\bibinfo {booktitle} {Proceedings of the 2024 SIAM International Conference
  on Data Mining (SDM)}}}\ (\bibinfo  {publisher} {SIAM},\ \bibinfo {year}
  {2024})\ pp.\ \bibinfo {pages} {490--498}\BibitemShut {NoStop}%
\bibitem [{\citenamefont {Kochkov}\ \emph {et~al.}(2021)\citenamefont
  {Kochkov}, \citenamefont {Pfaff}, \citenamefont {Sanchez-Gonzalez},
  \citenamefont {Battaglia},\ and\ \citenamefont {Clark}}]{kochkov2021GNN}%
  \BibitemOpen
  \bibfield  {author} {\bibinfo {author} {\bibfnamefont {D.}~\bibnamefont
  {Kochkov}}, \bibinfo {author} {\bibfnamefont {T.}~\bibnamefont {Pfaff}},
  \bibinfo {author} {\bibfnamefont {A.}~\bibnamefont {Sanchez-Gonzalez}},
  \bibinfo {author} {\bibfnamefont {P.}~\bibnamefont {Battaglia}},\ and\
  \bibinfo {author} {\bibfnamefont {B.~K.}\ \bibnamefont {Clark}},\ }\bibfield
  {title} {\bibinfo {title} {Learning ground states of quantum hamiltonians
  with graph networks},\ }\href {https://doi.org/10.48550/arXiv.2110.06390}
  {\bibfield  {journal} {\bibinfo  {journal} {arXiv:2110.06390}\ } (\bibinfo
  {year} {2021})}\BibitemShut {NoStop}%
\bibitem [{\citenamefont {Liu}\ \emph {et~al.}(2022)\citenamefont {Liu},
  \citenamefont {Gong}, \citenamefont {Li}, \citenamefont {Poilblanc},
  \citenamefont {Chen},\ and\ \citenamefont
  {Gu}}]{Zheng-ChengGu2022_TensorNetwork+PEPS}%
  \BibitemOpen
  \bibfield  {author} {\bibinfo {author} {\bibfnamefont {W.-Y.}\ \bibnamefont
  {Liu}}, \bibinfo {author} {\bibfnamefont {S.-S.}\ \bibnamefont {Gong}},
  \bibinfo {author} {\bibfnamefont {Y.-B.}\ \bibnamefont {Li}}, \bibinfo
  {author} {\bibfnamefont {D.}~\bibnamefont {Poilblanc}}, \bibinfo {author}
  {\bibfnamefont {W.-Q.}\ \bibnamefont {Chen}},\ and\ \bibinfo {author}
  {\bibfnamefont {Z.-C.}\ \bibnamefont {Gu}},\ }\bibfield  {title} {\bibinfo
  {title} {Gapless quantum spin liquid and global phase diagram of the spin-1/2
  $j_1$-$j_2$ square antiferromagnetic heisenberg model},\ }\href
  {https://doi.org/https://doi.org/10.1016/j.scib.2022.03.010} {\bibfield
  {journal} {\bibinfo  {journal} {Science Bulletin}\ }\textbf {\bibinfo
  {volume} {67}},\ \bibinfo {pages} {1034} (\bibinfo {year}
  {2022})}\BibitemShut {NoStop}%
\bibitem [{Note1()}]{Note1}%
  \BibitemOpen
  \bibinfo {note} {The main code and data for the aCNN are available on GitHub
  at https://github.com/rqHe1/aCNN.}\BibitemShut {Stop}%
\end{thebibliography}%

\end{document}